\documentclass[journal,twoside]{IEEEtran}

\usepackage{ifthen,xspace}

\usepackage[cmex10]{amsmath}
\usepackage{amsfonts,amssymb}
\usepackage{mathrsfs,cite}
\usepackage{graphicx,color,footmisc} %,pstool}
\usepackage{amsthm,algpseudocode}
\usepackage{afterpage}

\theoremstyle{plain}
\newtheorem{lemma}{Lemma}
\newtheorem{theorem}{Theorem}
\newtheorem{definition}{Definition}

\newtheorem{assumption}{Assumption}

\IEEEoverridecommandlockouts
\allowdisplaybreaks[1] 

\DeclareMathOperator*{\argmin}{arg~min}

\title{Marginal multi-Bernoulli filters: RFS derivation of MHT, JIPDA and association-based MeMBer}

\author{Jason~L.~Williams,~\IEEEmembership{Member,~IEEE}
\thanks{Manuscript received September 11, 2012; revised December 7, 2012; resubmitted August 15, 2013; revised May 22, 2014 and November 28, 2014; accepted December 1, 2014. The associate editor coordinating the review of this manuscript and approving it for publication was Prof. Ba-Ngu Vo.}%
\thanks{The author is with the National Security, Intelligence, Surveillance and Reconnaissance Division, Defence Science and Technology Organisation, Australia, and the School of Electrical and Electronic Engineering, University of Adelaide, Australia (e-mail: jason.williams@dsto.defence.gov.au).}}

\IEEEpubid{\copyright~2014 Crown}

\begin{document}
\maketitle
\begin{abstract}
Recent developments in random finite sets (RFSs) have yielded a variety of tracking methods that avoid data association. This paper derives a form of the full Bayes RFS filter and observes that data association is implicitly present, in a data structure similar to MHT. Subsequently, algorithms are obtained by approximating the distribution of associations. Two algorithms result: one nearly identical to JIPDA, and another related to the MeMBer filter. Both improve performance in challenging environments.
\end{abstract}

\begin{keywords}
random finite sets, conjugate prior, multiple hypothesis tracking, joint probabilistic data association, MeMBer, Poisson point process, multi-Bernoulli, loopy belief propagation
\end{keywords}

\section{Introduction}
\label{sec:Introduction}
{\noindent}Recently, much work has been devoted to random finite set (RFS)-based approximations such as the probability hypothesis density (PHD) \cite{Mah03,VoMa06}, cardinalised PHD (CPHD) \cite{Mah07b,VoVo07} and multiple target multi-Bernoulli (MeMBer) filters \cite{Mah07,VoVo09}. A key characteristic of these methods has been their tractability, which is the consequence of clever approximations that avoid data association, \textit{i.e.}\xspace, reasoning over the correspondence of measurements and targets. For example, the PHD approximates the posterior distribution as a Poisson point process (PPP), for which updates can be calculated without data association. The CPHD improves upon this by calculating the cardinality distribution, and using an independently, identically distributed (iid) cluster process approximation for which updates can be calculated efficiently; CPHD has been shown to improve performance significantly over PHD in challenging environments (\textit{e.g.}\xspace, \cite{VoVo07}). The MeMBer makes approximations in its derivation that avoid association and are reasoned to be of little consequence when the false alarm rate is low; these approximations introduced a significant cardinality bias that was corrected in the cardinality balanced MeMBer (CB-MeMBer) \cite{VoVo09}. 

Outside of the RFS field, two major approaches are joint probabilistic data association (JPDA) \cite{ForBar83} and multiple hypothesis tracking (MHT) \cite{Rei79}; both explicitly formulate and reason over association hypotheses, \textit{i.e.}\xspace, hypotheses for the correspondence of measurements and targets. JPDA seeks to marginalise over the random variables representing association in order to calculate the marginal distribution of each track; joint integrated PDA (JIPDA) \cite{MusEva04,ChaMor11} and joint integrated track splitting (JITS) \cite{MusEva09,ChaMor11} extend this by incorporating target existence as an additional random variable to be estimated. MHT seeks to calculate the most likely association hypothesis, returning a tracking estimate conditioned on the hypothesis. Track oriented MHT (TOMHT) \cite{Kur90} provides a tractable approach for implementing MHT by maintaining a hypothesis tree for each track, and finding the most likely global hypothesis indirectly (\textit{e.g.}\xspace, using Lagrangian relaxation \cite{PooGad06}).

\IEEEpubidadjcol

\subsection{Contributions}
\label{ss:Contributions}
This paper presents a derivation of the full Bayes RFS filter under commonly invoked assumptions (point measurements, PPP birth and false alarms), and proposes approximations that yield tractable solutions of the tracking problem. We observe that, while the RFS framework avoids the need for \emph{explicitly} modelling data association, it \emph{implicitly} arises in the full Bayes RFS filter. Contributions include:
\begin{itemize}
\item The derivation in Section~\ref{sec:Derivation} proves a conjugate prior form (\textit{i.e.}\xspace, a form that is preserved by prediction and update) for the full Bayes RFS filter in the model of interest. Various aspects of the form are similar to TOMHT (maintenance of a hypothesis tree for each track) and \cite{MorCho86} (use of a PPP to model targets which so-far remain undetected). In effect, the result provides an RFS-based derivation of the data structure employed by TOMHT, where the posterior involves a summation over association hypotheses that are similar to those utilised in MHT.
\item Observing that data association implicitly arises in the derivation, we subsequently seek to approximate the discrete distribution of data association. The first approximation yields the track oriented marginal MeMBer/Poisson (TOMB/P) filter, which is very similar to JITS and JIPDA (and related to the RFS-based derivation in \cite{MorCha03}); the differences relate to the inclusion of a Bayesian model of target birth in our RFS derivation. Thus, with minor changes, the popular JITS/JIPDA algorithms (and extensions thereof) can be derived within the RFS framework.
\item The second approximation yields the measurement oriented marginal MeMBer/Poisson (MOMB/P) filter, which collects all hypotheses using a given measurement into a single Bernoulli component in a manner similar to the MeMBer filter of \cite{Mah07,VoVo09}, but weighting hypotheses using the marginal association weights. Both TOMB/P and MOMB/P can be implemented using the high quality, tractable approximation of the marginal measurement-to-track association probabilities described in \cite{WilLau12}.
\item Section~\ref{sec:Results} demonstrates the proposed methods in a challenging scenario, showing that improved performance is obtained in comparison to CPHD and CB-MeMBer (particularly in cases with a lower probability of detection) for a similar computational load.
\end{itemize}

Preliminary versions of this paper were available in \cite{Wil11b,Wil11c,Wil14c}.

\section{Background}
\label{sec:Background}
RFS-based methods \cite{Mah07} have been developed in order to conduct statistical inference in problems in which the variables of interest and/or the observations form finite sets. The methods are particularly applicable to tracking since they address two of the major challenges involved:
\begin{itemize}
\item The number of targets present in the scene is unknown
\item Measurements are unordered, and measurement-to-target correspondence is unknown
\end{itemize} 
Problems such as this can be conveniently formulated by modelling the system state as a set of target states\footnote{We use the convention of capital letters (\textit{e.g.}\xspace, $X$) representing sets, and lower-case letters representing single members (\textit{e.g.}\xspace, $x$).} $X_t = \{x_t^1,\dots,x_t^{n_t}\}$ (at time $t$), and incorporating set-valued measurements $Z_t = \{z_t^1,\dots,z_t^{m_t}\}$. RFS densities encode both uncertainty in the number of targets, and their states. A RFS density can be constructed from a cardinality distribution $p(n)$, $n\geq 0$, and a series of cardinality-conditioned joint distributions $f_n(x_1,\dots,x_n)$, yielding
\begin{equation}
f(\{x_1,\dots,x_n\}) = p(n) \sum_{\pi} f_n(x_{\pi(1)},\dots,x_{\pi(n)})
\end{equation}
where the sum over $\pi$ spans the $n!$ permutation functions, ensuring that $f(X)$ is permutation invariant. Bayesian estimation can be performed in this framework (conceptually, at least) by interleaving prediction and update steps: \cite[p484]{Mah07}
\begin{align}
f_{t|t-1}(X) &= \int{f_{t|t-1}(X|X')f_{t-1|t-1}(X')\delta X'} \label{eq:RFSPrediction} \\
f_{t|t}(X) &= \frac{f(Z_t|X) f_{t|t-1}(X)}{f_{t|t-1}(Z_t)} \propto f(Z_t|X) f_{t|t-1}(X) \label{eq:RFSUpdate}
\end{align}
where $f_{t|t'}(X)$ is the RFS density at time $t$ given measurements up to and including time $t'$ (we leave off conditioning on the measurement set history $Z^{t'}=(Z_1,\dots,Z_{t'})$ throughout, as this is implicit in the second subscript $f_{t|t'}$), $f_{t|t-1}(X|X')$ is the RFS transition density, $f(Z|X)$ is the RFS measurement likelihood, and the set integral is defined as: \cite[p361]{Mah07}
\begin{equation}
\int{g(X) \delta X}
\triangleq g(\emptyset) + \sum_{n=1}^{\infty}\frac{1}{n!}\int{g(\{x_1,\dots,x_n\})\mathrm{d} x_1 \cdots \mathrm{d} x_n} \label{eq:SetIntegral}
\end{equation}
Probability generating functions (pgfs) are widely used in queueing systems, \textit{etc}\xspace, as they greatly simplify analysis of independent random variables (\textit{e.g.}\xspace, \cite{GriSti01}). Likewise, probability generating functionals\footnote{Note that, whereas the argument of a pgf $G(s)$ is the scalar $s$, the argument of a p.g.fl $G[h]$ is the function $h(x)$.} (p.g.fls) simplify analysis of RFSs that possess independence relationships. The p.g.fl of a RFS density $f$ is defined as: \cite[p371]{Mah07}
\begin{equation}
G[h] = \int{h^X f(X) \delta X}
\end{equation}
where $h^X \triangleq \prod_{x\in X}h(x)$. Just as the Fourier, Laplace and $z$ transforms are alternative representations of signals, a p.g.fl is an alternative representation of a RFS distribution. If $X$ and $Y$ are independent RFSs and $Z=X\cup Y$, then \cite[p372,386]{Mah07}
\begin{equation}\label{eq:Convolution}
\begin{split}
f_Z(Z) &= \sum_{X\subseteq Z}f_X(X) f_Y(Z-X)  \\
G_Z[h] &= G_X[h] G_Y[h] 
\end{split}
\end{equation}
This convolution identity is analogous to the result that the pgf of the sum of two independent variables $z=x+y$ is $G_z(s)=G_x(s) G_y(s)$ \cite[p153]{GriSti01}. 

This work utilises two basic forms of RFS distributions: PPPs and Bernoulli processes. A PPP with intensity function $\lambda(x)$ has RFS density and p.g.fl: \cite[p373]{Mah07}
\begin{equation}\label{eq:Poisson}
\begin{split}
f(X) &= \exp\left\{-\int{\lambda(x)\mathrm{d} x}\right\} \cdot \prod_{x\in X} \lambda(x)  \\
G[h] &= \exp\{\langle \lambda,h-1\rangle\} \propto \exp\{\langle \lambda,h\rangle\} 
\end{split}
\end{equation}
where $\langle \lambda,h\rangle$ denotes the inner product of $\lambda(\cdot)$ and $h(\cdot)$:
\begin{equation}
\langle \lambda,h\rangle \triangleq \int{\lambda(x)h(x)\mathrm{d} x}
\end{equation}
and thus $\langle \lambda,h-1\rangle = \int{\lambda(x)h(x)\mathrm{d} x} - \int{\lambda(x)\mathrm{d} x}$.

A Bernoulli process with probability of existence $r$ and existence-conditioned probability density function (PDF) $f(x)$ has RFS density and p.g.fl: 
\begin{equation}\label{eq:Bernoulli}
\begin{split}
f(X) &= \begin{cases}
1-r, & X = \emptyset \\
r \cdot f(x), & X = \{x\} \\
0, & \mbox{otherwise}
\end{cases}  \\
G[h] &= 1 - r + r \cdot \langle f,h\rangle 
\end{split}
\end{equation}
By (\ref{eq:Bernoulli}) and (\ref{eq:Convolution}), we can write the p.g.fl of a multi-Bernoulli process, \textit{i.e.}\xspace, the process resulting from the union $Y=\bigcup_{i=1}^N X_i$ of independent Bernoulli processes $X_i$:
\begin{equation}\label{eq:MultiBernoulliPGFL}
G_Y[h] = \prod_{i=1}^N \left( 1 - r_i + r_i\langle f_i,h \rangle\right)
\end{equation}
where $r_i$ and $f_i(x)$ are the existence probability and existence-conditioned PDF of the $i$-th Bernoulli process $X_i$. The RFS density of a multi-Bernoulli process is more difficult to deal with due to the sum over permutations. If $X=\{x_1,\dots,x_n\}$, \cite[p368]{Mah07} gives the multi-Bernoulli form as:
\begin{equation}
f(X) = \left[\prod_{i=1}^N(1-r_i)\right]\cdot\sum_{1\leq i_1\neq\cdots\neq i_n\leq N} \prod_{k=1}^n\left[\frac{r_{i_k}}{1-r_{i_k}}f_{i_k}(x_i) \right]
\end{equation}
An alternative form for this expression is:
\begin{equation}\label{eq:MultiBernoulli}
f(X) = \sum_{\alpha\in{\cal P}_N^n}
\prod_{i=1}^N f_i(X_{\alpha(i)}) 
\end{equation}
where $f_i(X)$ is the Bernoulli RFS density of the form (\ref{eq:Bernoulli}) with parameters $r_i$ and $f_i(x)$, and ${\cal P}_N^n$ is the set of all functions 
\ifCLASSOPTIONdraftcls
\begin{multline}\label{eq:DefinitionP}
{\cal P}_N^n = \big\{\alpha : \{1,\dots,N\} \rightarrow \{0,\dots,n\} \big| 
\{1,\dots,n\}\subseteq\alpha(\{1,\dots,N\}), \\
\mbox{ and if } \alpha(i)>0,\; i\neq j \mbox{ then } \alpha(i)\neq\alpha(j)
\big\}
\end{multline}
\else
\begin{multline}\label{eq:DefinitionP}
{\cal P}_N^n = \big\{\alpha : \{1,\dots,N\} \rightarrow \{0,\dots,n\} \big| \\
\{1,\dots,n\}\subseteq\alpha(\{1,\dots,N\}), \\
\mbox{ and if } \alpha(i)>0,\; i\neq j \mbox{ then } \alpha(i)\neq\alpha(j)
\big\}
\end{multline}
\fi
This is the set of permutation-like functions that map exactly one Bernoulli component index onto each $k\in\{1,\dots,n\}$, and the remaining Bernoulli component indices onto $0$. The sets $X_{\alpha(i)}$ are defined as:
\begin{equation}\label{eq:BernoulliSet}
X_{\alpha(i)} \triangleq \begin{cases}
\emptyset, & \alpha(i) = 0 \\
\{x_{\alpha(i)}\}, & \alpha(i)>0
\end{cases} 
\end{equation}

\subsection{Dynamics model and prediction step}
\label{ss:RFSPrediction}
Assumption \ref{ass:Dynamics} describes the dynamics model that we utilise in this work.
\begin{assumption}
The multiple target state evolves according to the following time dynamics process:
\begin{itemize}
\item Targets arrive at each time according to a non-homogeneous PPP with birth intensity $\lambda^\mathrm{b}(x)$, independent of existing targets
\item Targets depart according to iid Markovian processes; the survival probability in state $x$ is $P^{\mathrm{s}}(x)$
\item Target motion follows iid Markovian processes; the single-target transition PDF is $f_{t|t-1}(x|x')$
\end{itemize}
\label{ass:Dynamics}
\end{assumption}
Assumption \ref{ass:Dynamics} describes a multiple target dynamics process which is the union of a PPP (describing arrival of new targets), and an independent Bernoulli process for each existing target conditioning on the prior existence and state, $G[h|x'] = 1 - P^{\mathrm{s}}(x') + P^{\mathrm{s}}(x') p_h(x')$, where $p_h(x') \triangleq \int{h(x) f_{t|t-1}(x|x')\mathrm{d} x}$. Thus, the multiple target transition p.g.fl is: \cite[p474]{Mah07}
\ifCLASSOPTIONdraftcls
\begin{align}
G[h|X'] &= \int{h^X f_{t|t-1}(X|X') \delta X} \\
&= \exp\{\langle\lambda^\mathrm{b},h-1\rangle\} \prod_{x'\in X'}\{1 - P^{\mathrm{s}}(x') + P^{\mathrm{s}}(x') p_h(x')\} \label{eq:PGFLTransition}
\end{align}
\else
\begin{align}
\lefteqn{G[h|X'] = \int{h^X f_{t|t-1}(X|X') \delta X}} \\
&= \exp\{\langle\lambda^\mathrm{b},h-1\rangle\} \prod_{x'\in X'}\{1 - P^{\mathrm{s}}(x') + P^{\mathrm{s}}(x') p_h(x')\} \label{eq:PGFLTransition}
\end{align}
\fi
Taking the p.g.fl of (\ref{eq:RFSPrediction}), changing the order of integration and substituting (\ref{eq:PGFLTransition}), we obtain the p.g.fl form of the prediction equation: \cite[p529]{Mah07}
\ifCLASSOPTIONdraftcls
\begin{align}
G_{t|t-1}[h] &\triangleq \int{h^X f_{t|t-1}(X) \delta X} \\
&= \iint{h^X f_{t|t-1}(X|X')\delta X f_{t-1|t-1}(X') \delta X'} \\
&= \exp\{\langle\lambda^\mathrm{b},h-1\rangle\}\cdot \int{\{1-P^{\mathrm{s}} + P^{\mathrm{s}} p_h\}^{X'}f_{t-1|t-1}(X')\delta X'} \\
&= \exp\{\langle\lambda^\mathrm{b},h-1\rangle\} \cdot G_{t-1|t-1}[1-P^{\mathrm{s}} + P^{\mathrm{s}} p_h] \label{eq:PGFLPrediction}
\end{align}
\else
\begin{align}
\lefteqn{G_{t|t-1}[h] \triangleq \int{h^X f_{t|t-1}(X) \delta X}} \\
&= \iint{h^X f_{t|t-1}(X|X')\delta X f_{t-1|t-1}(X') \delta X'} \\
&= \exp\{\langle\lambda^\mathrm{b},h-1\rangle\}\cdot \int{\{1-P^{\mathrm{s}} + P^{\mathrm{s}} p_h\}^{X'}f_{t-1|t-1}(X')\delta X'} \\
&= \exp\{\langle\lambda^\mathrm{b},h-1\rangle\} \cdot G_{t-1|t-1}[1-P^{\mathrm{s}} + P^{\mathrm{s}} p_h] \label{eq:PGFLPrediction}
\end{align}
\fi

\subsection{Measurement update step and measurement model}
\label{ss:RFSUpdate}
In analogy to the similar result for pgfs \cite[p150]{GriSti01}, the RFS distribution can be recovered from the p.g.fl via the iterated functional derivative: \cite[p375-376,384]{Mah07}
\begin{equation}\label{eq:PGFLInversion}
f(X) = \left.\frac{\delta}{\delta X}G[h]\right|_{h=0} = \left.\frac{\delta^{|X|}}{\prod_{x\in X}\delta x}G[h]\right|_{h=0}
\end{equation}
where $\frac{\delta}{\delta x}G[h] \triangleq \lim_{\epsilon\downarrow 0}\frac{G[h + \epsilon\delta_x] - G[h]}{\epsilon}$, and $\delta_{x'}(x)$ is a Dirac delta function concentrated at $x=x'$.\footnote{A measure theoretic exploration of this can be found in \cite[Section 2.2.5]{Vo08}.} Thus we can recover the p.g.fl of the updated distribution (\ref{eq:RFSUpdate}) via \cite[p530]{Mah07}
\ifCLASSOPTIONdraftcls
\begin{align}
G_{t|t}[h] &\triangleq \int{h^X f_{t|t}(X)\delta X} \\
&\propto \int{h^X f(Z_t|X)f_{t|t-1}(X)\delta X} \\
&= \left.\frac{\delta}{\delta Z_t}\iint{h^X g^{Z_t} f(Z_t|X)f_{t|t-1}(X)\delta X\delta Z_t}\right|_{g=0} \\
&= \left.\frac{\delta}{\delta Z_t}
F[g,h]
\right|_{g=0} \label{eq:PGFLUpdate}
\end{align}
\else
\begin{align}
\lefteqn{G_{t|t}[h] \triangleq \int{h^X f_{t|t}(X)\delta X}} \\
&\propto \int{h^X f(Z_t|X)f_{t|t-1}(X)\delta X} \\
&= \left.\frac{\delta}{\delta Z_t}\iint{h^X g^{Z_t} f(Z_t|X)f_{t|t-1}(X)\delta X\delta Z_t}\right|_{g=0} \\
&= \left.\frac{\delta}{\delta Z_t}
F[g,h]
\right|_{g=0} \label{eq:PGFLUpdate}
\end{align}
\fi
where the derivative with respect to $Z_t$ operates on the corresponding functional $g$, and $F[g,h]$ is the joint p.g.fl of measurements $Z_t$ (functional $g$) and targets $X$ (functional $h$). Assumption \ref{ass:Measurement} describes the measurement model that we utilise in this work.
\begin{assumption}
The multiple target measurement process is as follows:
\begin{itemize}
\item Each target may give rise to at most one measurement; probability of detection in state $x$ is $P^{\mathrm{d}}(x)$
\item Each measurement is the result of at most one target
\item False alarm measurements arrive according to a non-homogeneous PPP with intensity $\lambda^\mathrm{fa}(z)$, independent of targets and target-related measurements
\item Each target-derived measurement is independent of all other targets and measurements conditioned on its corresponding target; the single target measurement likelihood is $f(z|x)$
\end{itemize}
\label{ass:Measurement}
\end{assumption}
In analogy with (\ref{eq:PGFLTransition}), the p.g.fl of the measurement likelihood $f(Z|X)$ is \cite[p422]{Mah07}
\ifCLASSOPTIONdraftcls
\begin{align}
G[g|X] &= \int{g^Z f(Z|X) \delta Z} \\
&= \exp\{\langle\lambda^\mathrm{fa},g-1\rangle\}\prod_{z\in Z}\{1-P^{\mathrm{d}}(x) + P^{\mathrm{d}}(x)p_g(x)\} \label{eq:PGFLMeasurement}
\end{align}
\else
\begin{align}
\lefteqn{G[g|X] = \int{g^Z f(Z|X) \delta Z}} \\
&= \exp\{\langle\lambda^\mathrm{fa},g-1\rangle\}\prod_{x\in X}\{1-P^{\mathrm{d}}(x) + P^{\mathrm{d}}(x)p_g(x)\} \label{eq:PGFLMeasurement}
\end{align}
\fi
where $p_g(x) \triangleq \int{g(z)f(z|x)\mathrm{d} z}$. Thus: \cite[p531]{Mah07}
\ifCLASSOPTIONdraftcls
\begin{align}
F[g,h] &= \int{h^X G[g|X] f_{t|t-1}(X)\delta X} \\
&= \exp\{\langle\lambda^\mathrm{fa},g-1\rangle\}\int{\{h(1-P^{\mathrm{d}} + P^{\mathrm{d}} p_g)\}^X f_{t|t-1}(X)\delta X} \\
&= \exp\{\langle\lambda^\mathrm{fa},g-1\rangle\} G_{t|t-1}[h(1-P^{\mathrm{d}} + P^{\mathrm{d}} p_g)] \label{eq:JointPGFL}
\end{align}
\else
\begin{align}
\lefteqn{F[g,h] = \int{h^X G[g|X] f_{t|t-1}(X)\delta X}} \\
&= \exp\{\langle\lambda^\mathrm{fa},g-1\rangle\}\int{\{h(1-P^{\mathrm{d}} + P^{\mathrm{d}} p_g)\}^X f_{t|t-1}(X)\delta X} \\
&= \exp\{\langle\lambda^\mathrm{fa},g-1\rangle\} G_{t|t-1}[h(1-P^{\mathrm{d}} + P^{\mathrm{d}} p_g)] \label{eq:JointPGFL}
\end{align}
\fi
Thus, using (\ref{eq:PGFLPrediction}), (\ref{eq:PGFLUpdate}) and (\ref{eq:JointPGFL}), a Bayes filter may be derived directly in p.g.fl form. A key component of the derivation is the product rule: \cite[p395]{Mah07}
\begin{equation}
\frac{\delta}{\delta Z}(F_0[h]\cdots F_n[h]) = \sum_{W_0 \uplus \cdots \uplus W_n = Z}
\frac{\delta F_0}{\delta W_0}[h] \cdots
\frac{\delta F_n}{\delta W_n}[h]
\label{eq:ProductRule}
\end{equation}
where the notation $\uplus$ denotes that the sum is over all disjoint sets $W_0,\dots,W_n$ such that $W_0 \cup \cdots \cup W_n=Z$. This permits calculation of the derivative of a product of several component p.g.fls (\textit{e.g.}\xspace, a multi-Bernoulli p.g.fl).

\section{Random set filter derivation}
\label{sec:Derivation}
The model of interest was detailed in Assumptions \ref{ass:Dynamics} and \ref{ass:Measurement}. The derivation does not change if the birth density, survival probability, detection probability, false alarm density and measurement likelihood are time varying (\textit{e.g.}\xspace, depending on a known, time-varying sensor state); we omit the time index from these parameters for notational simplicity. The multiple sensor case may be addressed by performing update steps for each sensor sequentially in between prediction steps.

Before we commence, we make several definitions that arise in the derivation, and define some of the elements that comprise the filter structure.
\begin{definition}\label{def:Unknown}
An \textbf{unknown target} is a target that is hypothesised to exist but has never been detected. 
\end{definition}
This terminology follows \cite{Rei79,MorCho86}. The explicit modelling of targets that have never been detected may appear unusual at first glance. For example, how can one distinguish targets that have never been detected from those that do not exist? Mathematically, it arises naturally: the RFS prediction model hypothesises birth of a PPP of targets (with intensity $\lambda^\mathrm{b}(x)$), and the detection model hypothesises that missed detections will occur on some proportion of these ($1-P^{\mathrm{d}}(x)$); thus we are obliged to carry them over to the next time step as \emph{unknown} targets. Practically, if $P^{\mathrm{d}}\approx 1$, the density of unknown targets will be very low and can be safely neglected. The practical necessity for maintaining awareness of unknown targets arises in cases in which sensor characteristics are non-uniform and non-stationary. For example, suppose that the sensor observes a region at irregular intervals, achieving a different probability of detection with each observation (as commonly occurs in phased array radar systems). In this case, even if the birth intensity is uniform and stationary, with each look one would expect to discover a different number of new targets; the unknown target density provides a mechanism for modelling this effect.

While measurement-to-target association is not explicitly introduced in the derivation, we will find that it arises implicitly, as a consequence of the sum over all decompositions of the measurements $Z_t$ into the disjoint subsets $W_i$ assigned to prior tracks (\textit{i.e.}\xspace, the factors of the p.g.fl of the prior), which appears in the product rule of (\ref{eq:ProductRule}). Thus we define:
\begin{definition}\label{def:GlobalHypothesis}
A \textbf{global association history hypothesis} (or \textbf{global hypothesis} for short) is a partitioning of all measurements received so far in to subsets, where each subset is hypothesised to correspond to a particular potential target. 
\end{definition}
Note that a consequence of assumption \ref{ass:Measurement} is that the subset of measurements for each target can contain at most one measurement from each time. Global association history hypotheses are made up of single target association history hypotheses, as defined below.

\begin{definition}\label{def:SingleTargetHypothesis}
A \textbf{single target association history hypothesis} (or \textbf{single target hypothesis} for short) is a subset of measurements that are hypothesised to correspond to the same potential target. 
\end{definition}
The term \emph{potential} target is used in the definitions above because single-target hypotheses correspond to Bernoulli distributions (i.e., the form (\ref{eq:Bernoulli}), incorporating a probability of existence and a PDF of target state). Consequently, a global hypothesis specifies a \emph{distribution} over target cardinality, rather than a unique cardinality. This enables a significantly more compact set of hypotheses than in conventional MHT developments. For example, one single target hypothesis can incorporate a distribution over the following events:
\begin{enumerate}
\item The target never existed
\item The target did exist but death occurred at some time since the last detection
\item The target continues to exists
\end{enumerate}
Incorporation of the first event arises from the derivation in theorem \ref{th:Update}, which shows that the update step naturally creates a new Bernoulli component for each measurement, where the probability of existence is the ratio of the PPP intensity of measurements arising from unknown targets to the sum of this intensity and the PPP intensity of false alarms, $\lambda^\mathrm{fa}(z)$. Thus, one single target hypothesis represents both the event that the corresponding measurement was false alarm, and the event that it was the first detection of a new target, via a single Bernoulli distribution.

A global hypothesis at time $t$ may be represented as $a=(a^1,\dots,a^n)$, where $a^i$ indexes the single target hypothesis utilised for the $i$-th target. We denote by ${\cal M}^{t'}$ the set of all measurement indices up to and including time $t'$, \textit{i.e.}\xspace, the elements of ${\cal M}^{t'}$ are of the form $(\tau,j)$, where $j\in\{1,\dots,m_{\tau}\}$ is an index of a measurement in scan $\tau\leq t'$, where $t'$ denotes the last scan of measurements incorporated into the filter, \textit{e.g.}\xspace, $t'=t-1$ following the prediction step, and $t'=t$ following the update step. Each single-target association history hypothesis incorporates the following information:
\begin{itemize}
\item the history of measurement indices that are hypothesised to correspond to the target under the hypothesis, ${\cal M}^{t'}(i,a^i)\subseteq{\cal M}^{t'}$ (\textit{e.g.}\xspace, if ${\cal M}^{t'}(i,a^i)=\{(3,7),(5,8)\}$ at time $t'=6$, then $a^i$ hypothesises that the $i$-th hypothesised target was first detected as measurement index $7$ at time $3$, a missed detection occurred at time $4$, it was detected as measurement $8$ at time $5$, and a missed detection occurred again at time $6$)
\item the hypothesis weight $w_{t|t'}^{i,a^i}$ (utilised in calculation of the probability of the global hypotheses)
\item the hypothesis-conditioned Bernoulli distribution $f_{t|t'}^{i,a^i}(X)$ (with p.g.fl $G_{t|t'}^{i,a^i}[h]$) of the form (\ref{eq:Bernoulli}), parameterised by the probability of existence under the hypothesis, $r_{t|t'}^{i,a^i}$, and the PDF under the hypothesis, $f_{t|t'}^{i,a^i}(x)$
\end{itemize}

Single-target hypotheses are grouped together into \emph{tracks}, where the point in common is that all hypothesise the same measurement corresponding to the first detection of the target.\footnote{For each track, there is also a hypothesis that the target \emph{never} existed, and thus uses no measurements.}

\begin{definition}\label{def:Track}
A \textbf{track} is the collection of information available about the target that was first detected in a particular measurement, consisting of a collection of single-target hypotheses (and the accompanying information) representing different possibilities of measurement sequences corresponding to the target.
\end{definition}
Note that this definition is more in line with the use of the term in JPDA and related methods than MHT.\footnote{The term \emph{track} in \cite{Kur90} refers to the single-target hypothesis of definition \ref{def:SingleTargetHypothesis}, while the \emph{target tree}, or the set of \emph{track hypotheses} in \cite{Kur90} refers to the track of definition \ref{def:Track}.} As a consequence of definition \ref{def:Track}, a new track is created for each measurement received (conceptually, at least). The structure of global hypotheses, single-target hypotheses and tracks is similar to that in the TOMHT \cite{Kur90}; it is illustrated in Fig.~\ref{fig:MBMStructure}. The advantage of the structure is the compactness of representation that it achieves, since the same single-target hypotheses are utilised in many global hypotheses.

We denote by ${\cal T}_{t|t'}=\{1,\dots,n_{t|t'}\}$ the tracks at time $t$ conditioned on measurements up to and including time $t'$, and by ${\cal H}^i_{t|t'}=\{1,\dots,h^i_{t|t'}\}$ the single-target hypotheses in track $i\in{\cal T}_{t|t'}$.

Our derivation is based on an induction process, assuming a particular form for the full multiple target probability distribution, and showing that this form is maintained by prediction and update steps. The form of the full multi-target distribution at time $t$, conditioned on measurements up to and including time $t'$ (\textit{i.e.}\xspace, if $t'=t-1$ then $f_{t|t'}(X_t)=f(X_t|Z^{t-1})$ is the predicted distribution, and if $t'=t$ then $f_{t|t'}(X_t)=f(X_t|Z^t)$ is the updated distribution), consists of two independent components:
\begin{equation}
\begin{split}
f_{t|t'}(X) &= \sum_{Y\subseteq X}f^\mathrm{ppp}_{t|t'}(Y) f^\mathrm{mbm}_{t|t'}(X-Y) \\
G_{t|t'}[h] &= G^\mathrm{ppp}_{t|t'}[h] G^\mathrm{mbm}_{t|t'}[h]
\end{split}
\label{eq:FullDensity}
\end{equation}
where $f^\mathrm{ppp}_{t|t'}(X)$ (and the corresponding p.g.fl $G^\mathrm{ppp}_{t|t'}[h]$) is a PPP representing unknown targets, with intensity $\lambda^u_{t|t'}(x)$:
\begin{equation}
\begin{split}
f^\mathrm{ppp}_{t|t'}(X) &= \exp\{-\langle\lambda^u_{t|t'},1\rangle\}\prod_{x\in X}\lambda^u_{t|t'}(x) \\
G^\mathrm{ppp}_{t|t'}[h] &= \exp\{\langle\lambda^u_{t|t'},h-1\rangle\}
\end{split}
\label{eq:UndetectedTargets}
\end{equation}
and $f^\mathrm{mbm}_{t|t'}(X)$ (and the corresponding p.g.fl $G^\mathrm{mbm}_{t|t'}[h]$) is a multi-Bernoulli mixture (MBM), \textit{i.e.}\xspace, a linear combination of multi-Bernoulli distributions:
\begin{align}
f^\mathrm{mbm}_{t|t'}(\{x_1,\dots,x_n\}) &= \sum_{\alpha\in{\cal P}_{n_{t|t'}}^n,a\in{\cal A}^{t|t'}} w^a
\prod_{i\in{\cal T}_{t|t'}}f_{t|t'}^{i,a^i}(X_{\alpha(i)}) \notag\\
G^\mathrm{mbm}_{t|t'}[h] &= \sum_{a\in{\cal A}^{t|t'}} w^a\prod_{i\in{\cal T}_{t|t'}}G_{t|t'}^{i,a^i}[h]
\label{eq:PreexistingTargets}
\end{align}
where ${\cal P}_{n_{t|t'}}^n$ and $X_{\alpha(i)}$ were defined in (\ref{eq:DefinitionP}) and (\ref{eq:BernoulliSet}), and
${\cal A}^{t|t'}$ is the set of global association history hypotheses. Each global hypothesis must incorporate a single-target hypothesis for each track, and each global hypothesis must explain the origin of each measurement (up to and including time $t'$), thus
\ifCLASSOPTIONdraftcls
\begin{equation}\label{eq:AssocAlphabet}
{\cal A}^{t|t'} = \bigg\{
(a^1,\dots,a^{n_{t|t'}})
\bigg| 
a^i\in{\cal H}^i_{t|t'}, 
\bigcup_{i\in{\cal T}_{t|t'}}{\cal M}^{t'}(i,a^i)={\cal M}^{t'},
{\cal M}^{t'}(i,a^i)\cap{\cal M}^{t'}(j,a^j)=\emptyset \;\forall\; i\neq j 
\bigg\}
\end{equation}
\else
\begin{multline}\label{eq:AssocAlphabet}
{\cal A}^{t|t'} = \bigg\{
(a^1,\dots,a^{n_{t|t'}})
\bigg| 
a^i\in{\cal H}^i_{t|t'}, 
\bigcup_{i\in{\cal T}_{t|t'}}{\cal M}^{t'}(i,a^i)={\cal M}^{t'},
\\
{\cal M}^{t'}(i,a^i)\cap{\cal M}^{t'}(j,a^j)=\emptyset \;\forall\; i\neq j 
\bigg\}
\end{multline}
\fi
The global hypothesis weights $w^a$ are related to the single-target hypothesis weights via the expression
\begin{equation}\label{eq:GlobalHypWeight}
w^a \propto \prod_{i\in{\cal T}_{t|t'}}w_{t|t'}^{i,a^i}
\end{equation}
where the proportionality denotes that normalisation is required to ensure that $\sum_{a\in{\cal A}^{t|t'}}w^a=1$.

\subsection{Prediction step}
Theorem~\ref{th:Prediction} derives the prediction step by substituting (\ref{eq:FullDensity}) into (\ref{eq:PGFLPrediction}). The theorem confirms what one would intuitively expect: that the prediction of the PPP component (\ref{eq:UndetectedPropagation}) follows the standard PHD prediction step of \cite{Mah03,VoMa06}, and that the multi-Bernoulli tracks are predicted independently in an equivalent manner to the MeMBer \cite{Mah07,VoVo09} (excluding birth of new targets, since we utilise a PPP birth model as opposed to a multi-Bernoulli birth model).

\begin{theorem}\label{th:Prediction}
Assume that the distribution from the previous time step $G_{t-1|t-1}[h]$ is of the form given in (\ref{eq:FullDensity})-(\ref{eq:GlobalHypWeight}). Then the predicted distribution for the next step $G_{t|t-1}[h]$ is of the same form, with:
{\allowdisplaybreaks \begin{align}
\lambda^u_{t|t-1}(x) &= \lambda^\mathrm{b}(x) + \int{f_{t|t-1}(x|x')P^{\mathrm{s}}(x')\lambda^u_{t-1|t-1}(x')\mathrm{d} x'}
\label{eq:UndetectedPropagation} \\
n_{t|t-1} &= n_{t-1|t-1}; \quad h^i_{t|t-1} = h^i_{t-1|t-1} \; \forall \; i \\
w_{t|t-1}^{i,a^i} &= w_{t-1|t-1}^{i,a^i} \; \forall \; i, a^i \\
r_{t|t-1}^{i,a^i} &= r_{t-1|t-1}^{i,a^i}\langle f_{t-1|t-1}^{i,a^i},P^{\mathrm{s}}\rangle \; \forall \; i, a^i \label{eq:ExistProbPropagation} \\
f_{t|t-1}^{i,a^i}(x) &= \frac{\int{f_{t|t-1}(x|x')P^{\mathrm{s}}(x')f_{t-1|t-1}^{i,a^i}(x')\mathrm{d} x'}}{\langle f_{t-1|t-1}^{i,a^i},P^{\mathrm{s}}\rangle}\; \forall \; i, a^i \label{eq:KinematicDistPropagation}
\end{align}}
\end{theorem}
The proof of the theorem is in Appendix~\ref{app:PredictionProof}.

\vspace*{-9pt}

\subsection{Measurement update step}
\label{ss:UpdateDerivation}
\begin{figure}[tb]
\centering
\includegraphics[width=3.3in]{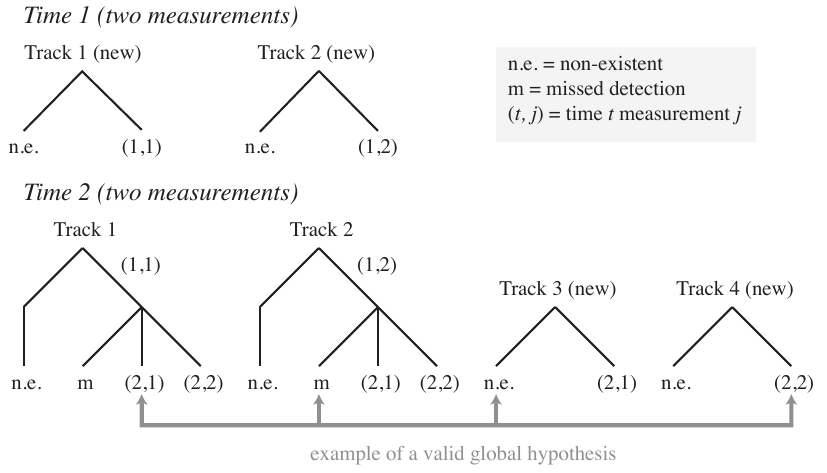} %
\caption{Tracks and hypotheses maintained by filter. Structure after time $1$ is shown at the top (assuming that there were two measurements). A new track is created for each measurement; this results from updating the PPP of unknown targets with each measurement. The new tracks each contain two hypotheses; one hypothesising that the measurement goes with a previously existing track and hence that the new track is not required (following eq (\ref{eq:NewTargetNonExistWQ})), and the other hypothesising that the measurement goes with the new track, capturing both the possibility that it is the result of a target detected for the first time, or a false alarm (following eq (\ref{eq:PoisUpdateMeasSet})-(\ref{eq:PoisUpdateKin})). The structure after time $2$ is shown at the bottom (assuming that there were also two measurements in the new scan). Each track from the prior distribution is continued, incorporating a hypothesis for each prior hypothesis (marked with ``m'', corresponding to a missed detection, following eq (\ref{eq:MissUpdateMeasSet})-(\ref{eq:MissUpdateKin})), and for each combination of a prior hypothesis and a new measurement (marked with ``($t,j)$'', corresponding to the prior hypothesis being updated with the measurement, following eq (\ref{eq:DetUpdateMeasSet})-(\ref{eq:DetUpdateKin})). Non-existence hypotheses are continued without branching. A branching structure similar to TOMHT results. New tracks are created for each new measurement (marked with ``new''), following the structure described for the first time step; again, these correspond to updating the PPP of unknown targets with each new measurement. The gray line shows an example of a global hypothesis---\textit{i.e.}\xspace, a choice of one single-target hypothesis from the tree for each track, in which every measurement in every scan is used exactly once. Each term in the sum in (\ref{eq:PreexistingTargets}) corresponds to a global hypothesis.
}\label{fig:MBMStructure} \vspace*{-9pt}
\end{figure}

Theorem~\ref{th:Update} derives the measurement update step by substituting (\ref{eq:JointPGFL}) and (\ref{eq:FullDensity}) into (\ref{eq:PGFLUpdate}). The PPP intensity update equation is (\ref{eq:UndetectedUpdate}); this is identical to a PHD update with no measurements \cite{Mah03,VoMa06}. The MBM update is illustrated and described in Fig.~\ref{fig:MBMStructure}.

\begin{theorem}\label{th:Update}
Assume that the predicted distribution $G_{t|t-1}[h]$ is of the form given in (\ref{eq:FullDensity})-(\ref{eq:GlobalHypWeight}). Then the updated distribution $G_{t|t}[h]$ (updated with the measurement set $Z_t=\{z_t^1,\dots,z_t^{m_t}\}$) is of the same form, with $n_{t|t} = n_{t|t-1} + m_t$, 
{\allowdisplaybreaks \begin{align}
\lambda^u_{t|t}(x) &= \{1-P^{\mathrm{d}}(x)\}\lambda^u_{t|t-1}(x) \label{eq:UndetectedUpdate} \\
{\cal M}^t &= {\cal M}^{t-1} \cup \big\{(t,j)|j\in\{1,\dots,m_t\}\big\} 
\end{align}}
For tracks continuing from previous time steps ($i\in\{1,\dots,n_{t|t-1}\}$), a hypothesis is included for each combination of a hypothesis from a previous time, and either a missed detection, or an update using one of the $m_t$ new measurements, such that the number of hypotheses becomes $h^i_{t|t} = h^i_{t|t-1}(1+m_t)$. For missed detection hypotheses ($i\in\{1,\dots,n_{t|t-1}\}$, $a^i\in\{1,\dots,h_{t|t-1}\})$:
\begin{align}
{\cal M}^t(i,a^i) &= {\cal M}^{t-1}(i,a^i) \label{eq:MissUpdateMeasSet} \\
w_{t|t}^{i,a^i} &= w^{i,a^i}_{t|t-1}(1-r_{t|t-1}^{i,a^i}+r_{t|t-1}^{i,a^i}\langle f_{t|t-1}^{i,a^i},1-P^{\mathrm{d}}\rangle) \label{eq:MissUpdateW} \\
r^{i,a^i}_{t|t} &= \frac{r_{t|t-1}^{i,a^i}\langle f_{t|t-1}^{i,a^i},1-P^{\mathrm{d}}\rangle}{1-r_{t|t-1}^{i,a^i}+r_{t|t-1}^{i,a^i}\langle f_{t|t-1}^{i,a^i},1-P^{\mathrm{d}}\rangle} \label{eq:MissUpdatePex}\\
f^{i,a^i}_{t|t}(x) &= \frac{\{1-P^{\mathrm{d}}(x)\}f_{t|t-1}^{i,a^i}(x)}{\langle f_{t|t-1}^{i,a^i},1-P^{\mathrm{d}}\rangle} \label{eq:MissUpdateKin}
\end{align}
For hypotheses updating existing tracks ($i\in\{1,\dots,n_{t|t-1}\}$, $a^i=\tilde{a}^i+h^i_{t|t-1} j$, $\tilde{a}^i\in\{1,\dots,h^i_{t|t-1}\}$, $j\in\{1,\dots,m_t\}$, \textit{i.e.}\xspace, the previous hypothesis $\tilde{a}^i$, updated with measurement $z_t^j$):\footnote{A hypothesis at the previous time with $r^{i,a^i}_{t|t-1}=0$ need not be updated since the posterior weight in (\ref{eq:DetUpdateW}) would be zero. For simplicity, the hypothesis numbering does not account for this exclusion.}
\begin{align}
{\cal M}^t(i,a^i) &= {\cal M}^{t-1}(i,\tilde{a}^i) \cup \{(t,j)\} \label{eq:DetUpdateMeasSet} \\
w_{t|t}^{i,a^i} &= w^{i,\tilde{a}^i}_{t|t-1}r_{t|t-1}^{i,\tilde{a}^i}\langle f_{t|t-1}^{i,\tilde{a}^i},f(z_t^j|\cdot)P^{\mathrm{d}}\rangle \label{eq:DetUpdateW}\\
r^{i,a^i}_{t|t} &= 1 \label{eq:DetUpdatePex}\\
f^{i,a^i}_{t|t}(x) &= \frac{f(z_t^j|x)P^{\mathrm{d}}(x)f_{t|t-1}^{i,\tilde{a}^i}(x)}{\langle f_{t|t-1}^{i,\tilde{a}^i},f(z_t^j|\cdot)P^{\mathrm{d}}\rangle} \label{eq:DetUpdateKin}
\end{align}
Finally, for new tracks, $i\in\{n_{t|t-1}+j\}$, $j\in\{1,\dots,m_t\}$ (\textit{i.e.}\xspace, the new track commencing on measurement $z_t^j$),
\begin{align}
h^i_{t|t} &= 2 \\
{\cal M}^t(i,1) &= \emptyset, \quad
w^{i,1}_{t|t} = 1, \quad r^{i,1}_{t|t} = 0 \label{eq:NewTargetNonExistWQ} \\
{\cal M}^t(i,2) &= \{(t,j)\} \label{eq:PoisUpdateMeasSet} \\
w^{i,2}_{t|t} &=  \lambda^\mathrm{fa}(z_t^j) + \langle\lambda^u_{t|t-1},f(z_t^j|\cdot)P^{\mathrm{d}}\rangle \label{eq:PoisUpdateW} \\
r_{t|t}^{i,2} &= \frac{\langle\lambda^u_{t|t-1},f(z_t^j|\cdot)P^{\mathrm{d}}\rangle}{\lambda^\mathrm{fa}(z_t^j) + \langle\lambda^u_{t|t-1},f(z_t^j|\cdot)P^{\mathrm{d}}\rangle} \label{eq:PoisUpdatePex}\\
f_{t|t}^{i,2}(x) &= \frac{f(z_t^j|x)P^{\mathrm{d}}(x)\lambda^u_{t|t-1}(x)}{\langle\lambda^u_{t|t-1},f(z_t^j|\cdot)P^{\mathrm{d}}\rangle} \label{eq:PoisUpdateKin} 
\end{align}
\end{theorem}
Note that from the last section of the theorem, a new track is created for each measurement. These tracks contain two single target hypotheses ($h^i_{t|t}=2$), the first of which covers the case that the measurement is associated with another track, hence the new track has probability of existence equal to zero (the PDF $f_{t|t}^{i,1}(x)$ has no effect and has not been specified); and the second of which covers the case that the measurement is not associated with any previous track, hence it is either a false alarm or a new target (the probability of existence $r_{t|t}^{i,2}$ models the relative likelihood of these two events). 

The proof of Theorem~\ref{th:Update} is in Appendix \ref{app:UpdateProof}\xspace. The first part of the proof (Lemma~\ref{lem:Structure}) shows that, in the updated distribution, the portion of the component $G^\mathrm{ppp}_{t|t-1}[h]$ that is not detected is independent (in the posterior) of the remainder of the distribution (which comprises the portion of $G^\mathrm{ppp}_{t|t-1}[h]$ that is detected, and both detected and undetected portions of the component $G^\mathrm{mbm}_{t|t-1}[h]$). Note that this is \emph{not} claiming that the \emph{entire} undetected portion is independent of the detected portion.

The updates in (\ref{eq:MissUpdateKin}), (\ref{eq:DetUpdateKin}) and (\ref{eq:PoisUpdateKin}) are standard single target measurement updates, and thus can be calculated with well-known methods (\textit{e.g.}\xspace, \cite{RisAru04,JulUhl04}). The derivation incorporates hypotheses updating every prior hypothesis with every measurement; however, in practical implementations, gating can be used to reduce the computational burden by excluding hypotheses with negligible weights. Other standard approximations such as clustering, mixture reduction, \textit{etc}\xspace are also required in practice.

\subsection{Initialisation}
One significant benefit of the inclusion of a Poisson component is in initialisation of the tracker. Generally, when a surveillance system is first activated, the fact that prior measurements are not available does not imply that no targets are present. The Poisson distribution provides a convenient mechanism for specifying a prior distribution on the number and position of targets when little information is available. By initialising $n_{0|0}=0$ (\textit{i.e.}\xspace, no MBM components), and setting $\lambda^u_{0|0}(x)=\lambda^u_0 f_0(x)$ where $\lambda^u_0$ is the expected number of targets present, and $f_0(x)$ is the (presumably diffuse) prior state distribution, the Bayes prediction and update steps incorporate this prior knowledge, adjusting the accrual of confidence in the existence probabilities $r^{i,a^i}_{t|t}$ accordingly. In \cite{Wil12F2}, it was demonstrated that this approach can considerably improve the speed of track initiation in low signal-to-noise ratio (SNR) environments over methods that do not utilise such prior knowledge.

\section{Marginal association filters}
\label{sec:Filters}
The previous section derived a conjugate prior form for the full Bayes RFS filter. Not surprisingly, the filter is intractable, and approximations must be made. Noting its similarity to the PHD, the PPP component of the posterior is tractable (\textit{e.g.}\xspace, see \cite{Wil12F2}). The difficulty is the number of global association history hypotheses, each of which gives rise to a term in the sum in the MBM (\ref{eq:PreexistingTargets}). To motivate our approximation, consider the definition of a probability of joint association history:
\begin{equation}\label{eq:JointAssocProb}
P_{t|t}(a) = w^a \propto \prod_{i\in{\cal T}_{t|t}}w_{t|t}^{i,a^i}
\end{equation}
following which (\ref{eq:PreexistingTargets}) can be equivalently viewed as a total probability expansion over this random variable
\begin{equation}
f^\mathrm{mbm}_{t|t}(X) = \sum_{\alpha\in{\cal P}_{n_{t|t}}^{|X|},a\in{\cal A}^{t|t}} P_{t|t}(a) \prod_{i\in{\cal T}_{t|t}}f_{t|t}^{i,a^i}(X_{\alpha(i)}) 
\label{eq:LCMBIndepApprox}
\end{equation}
The methods in this section approximate the posterior distribution $f^\mathrm{mbm}_{t|t}(X)$ by directly approximating the discrete probability distribution of global association history hypotheses $P_{t|t}(a)$. Herein we will assume that the prior distribution $f^\mathrm{mbm}_{t-1|t-1}(X)$ is multi-Bernoulli, such that the predicted distribution $f^\mathrm{mbm}_{t|t-1}(X)$ is multi-Bernoulli, and our task is to gain an approximation of $f^\mathrm{mbm}_{t|t}(X)$ that is multi-Bernoulli (since the true distribution is not). The predicted distribution in the form (\ref{eq:PreexistingTargets}) will be multi-Bernoulli if there is a single association hypothesis, \textit{e.g.}\xspace, 
\begin{equation}\label{eq:MultiBernoulliAssumption}
\renewcommand{\arraycolsep}{1pt}
\begin{array}{rlrl}
{\cal M}^{t-1} &= \emptyset, & h_{t|t-1}^i &= 1\;\forall\;i, \\
{\cal M}^{t-1}(i,a^i) &= \emptyset\;\forall\;i,a^i,\quad & w_{t|t-1}^{i,a^i} &= 1\;\forall\;i,a^i 
\end{array}
\end{equation}
In this case, we refer to the weight, existence probability and PDF for track $i$ as $w^i_{t|t}=1$, $r^i_{t|t}$ and $f^i_{t|t}(x)$ respectively.\footnote{We will see that $f^i_{t|t}(x)$ is a marginal distribution averaging over association hypotheses. The importance of there being a single association hypothesis is that constraints between hypotheses in different tracks are discarded.}

\subsection{Track-oriented marginal MeMBer-Poisson filter}
\label{ss:TOMB}
{\noindent}Suppose we seek an approximate representation of the distribution $P_{t|t}(a)$ in which the hypotheses for each track $a^i$ comprising the global hypothesis $a=(a^1,\dots,a^{n_{t|t}})$ are forced to be independent, \textit{i.e.}\xspace, a distribution of the form:
\begin{equation}\label{eq:MargTrackFilterWeights}
P_{t|t}(a) \approx \prod_{i\in{\cal T}_{t|t}} q^i(a^i)
\end{equation}
where we effectively expand the alphabet of the discrete random variable $a$ from ${\cal A}^{t|t}$ defined in (\ref{eq:AssocAlphabet}) to the Cartesian product $\tilde{{\cal A}}^{t|t}\triangleq\prod_i{\cal H}^i_{t|t}$, defining $P_{t|t}(a)=0$ for infeasible association events $a\notin{\cal A}^{t|t}$. Then with no further approximation, the joint multi-target distribution in (\ref{eq:LCMBIndepApprox}) becomes:
\begin{align}
f^\mathrm{mbm}_{t|t}(X) &\approx \sum_{\alpha\in{\cal P}_{n_{t|t}}^{|X|},a\in\tilde{{\cal A}}^{t|t}} \left[\prod_{i\in{\cal T}_{t|t}} q^i(a^i)\right] \cdot \prod_{i\in{\cal T}_{t|t}}f_{t|t}^{i,a^i}(X_{\alpha(i)})  \\
&= \sum_{\alpha\in{\cal P}_{n_{t|t}}^{|X|},a\in\tilde{{\cal A}}^{t|t}} \prod_{i\in{\cal T}_{t|t}}q^i(a^i)f_{t|t}^{i,a^i}(X_{\alpha(i)})  \\
&= \sum_{\alpha\in{\cal P}_{n_{t|t}}^{|X|}}\prod_{i\in{\cal T}_{t|t}}f_{t|t}^i(X_{\alpha(i)}) \label{eq:MargTrackFilter}\\
f_{t|t}^i(X) &= \sum_{a^i\in{\cal H}_{t|t}^i}q^i(a^i)f_{t|t}^{i,a^i}(X)
\end{align}
where (\ref{eq:MargTrackFilter}) exploits separability of the sum over $a\in\tilde{{\cal A}}_{t|t}$, e.g., that $\sum_{i\in{\cal I},j\in{\cal J}} f(i)g(j)=\big(\sum_{i\in{\cal I}}f(i)\big)\big(\sum_{j\in{\cal J}}g(j)\big)$. Comparing (\ref{eq:MargTrackFilter}) with (\ref{eq:MultiBernoulli}), we see that the resulting distribution is multi-Bernoulli. 

The problem of finding the choice of $q^i(a^i)$ which best fits the assumed form of the distribution can be solved by minimising the Kullback-Leibler (KL) divergence:
\begin{equation}\label{eq:TOMKLOpt}
\left[q^{i*}(a^i)\right]_{i\in{\cal T}_{t|t}} = \argmin_{q^i(a^i),\,i\in{\cal T}_{t|t}}D\bigg(P_{t|t}(a)\bigg|\bigg|\prod_{i\in{\cal T}_{t|t}}q^i(a^i) \bigg)
\end{equation}
where $D(p||q) = \sum_a p(a)\log\frac{p(a)}{q(a)}$. It is shown in \cite[p277]{KolFri09} that $q^{i*}(a^i) =  P_{t|t}^i(a^i) = \sum_{\tilde{a}|\tilde{a}^i=a^i}P_{t|t}(\tilde{a})$, \textit{i.e.}\xspace, the marginal association distribution for track $i$. 

Accordingly, the data carried forward to the next time for the MBM component are simply the single hypothesis Bernoulli distributions $f_{t|t}^i(X)$ (parameterised by an existence probability, and a position distribution consisting of a weighted mixture of the old hypothesis-conditioned distributions). We refer to this as the track oriented marginal MeMBer/Poisson filter (TOMB/P), as it forms tracks out of the marginal track-to-measurement association distributions. 
The approximation made in the TOMB/P is exactly that made in JPDA: approximating the joint association distribution by the product of its marginals.\footnote{JPDA/JIPDA additionally approximates the posterior distribution of each target as being Gaussian.\label{fn:JPDAGaussian}} Whereas JPDA assumes target existence, applying the marginal approximation within the RFS framework yields an algorithm that naturally captures target existence. 
In what follows, we show that the TOMB/P is equivalent to a particular variant of JITS/JIPDA\footref{fn:JPDAGaussian} with two minor modifications. The specific variant is JITS/JIPDA with a parametric clutter model, and using the one-point track initialisation method incorporating a new object spatial density \cite[p324]{ChaMor11}. We assume that $P^{\mathrm{d}}(x)=P^{\mathrm{d}}$ and $P^{\mathrm{s}}(x)=P^{\mathrm{s}}$ (\textit{i.e.}\xspace, that both of these parameters are spatially constant). From (\ref{eq:JointAssocProb}), (\ref{eq:MissUpdateW}), (\ref{eq:DetUpdateW}), (\ref{eq:PoisUpdateW}) and (\ref{eq:MultiBernoulliAssumption}), the probability of a global association hypothesis $a$ is:
\ifCLASSOPTIONdraftcls
\begin{align}
P_{t|t}(a) \propto& \prod_{i=1}^{n_{t|t}} w^{i,a^i}_{t|t} \notag \\
=& \prod_{i\in\{1,\dots,n_{t|t-1}\}|a^i=0} \left\{ 1- r^i_{t|t-1}P^{\mathrm{d}} \right\} 
\cdot \prod_{i\in\{1,\dots,n_{t|t-1}\}|a^i>0} \left\{r^i_{t|t-1}P^{\mathrm{d}} \langle f^i_{t|t-1},f(z_t^{a^i}|\cdot)\rangle \right\} \notag\\
& \times \prod_{j\in\{1,\dots,m_t\}|a^{n_{t|t-1}+j}=2}\left\{\lambda^\mathrm{fa}(z_t^j)+P^{\mathrm{d}}\langle\lambda^u_{t|t-1},f(z_t^j|\cdot)\rangle\right\} \label{eq:JointEventProb1}
\end{align}
\else
\begin{align}
P_{t|t}&(a) \propto \prod_{i=1}^{n_{t|t}} w^{i,a^i}_{t|t} \notag \\
=& \prod_{i\in\{1,\dots,n_{t|t-1}\}|a^i=0} \left\{ 1- r^i_{t|t-1}P^{\mathrm{d}} \right\}  \notag\\
& \times \prod_{i\in\{1,\dots,n_{t|t-1}\}|a^i>0} \left\{r^i_{t|t-1}P^{\mathrm{d}} \langle f^i_{t|t-1},f(z_t^{a^i}|\cdot)\rangle \right\} \notag\\
& \times \prod_{j\in\{1,\dots,m_t\}|a^{n_{t|t-1}+j}=2}\left\{\lambda^\mathrm{fa}(z_t^j)+P^{\mathrm{d}}\langle\lambda^u_{t|t-1},f(z_t^j|\cdot)\rangle\right\} \label{eq:JointEventProb1}
\end{align}
\fi
Since, under any hypothesis, a measurement will either be explained as belonging to an existing track ($\exists \; i \mbox{ s.t.\ } a^i=j$), or a new track ($a^{n_{t|t-1}+j}=2$) (which incorporates the possibility of it being a false alarm), we can divide (\ref{eq:JointEventProb1}) by the constant $\prod_{j=1}^{m_t}\left\{\lambda^\mathrm{fa}(z_t^j)+P^{\mathrm{d}}\langle\lambda^u_{t|t-1},f(z_t^j|\cdot)\rangle\right\}$ to obtain:
\ifCLASSOPTIONdraftcls
\begin{align}
P_{t|t}(a) \propto& \prod_{i\in\{1,\dots,n_{t|t-1}\}|a^i=0} \left\{ 1- r^i_{t|t-1}P^{\mathrm{d}} \right\} 
\cdot \prod_{i\in\{1,\dots,n_{t|t-1}\}|a^i>0} \left\{\frac{
r^i_{t|t-1}P^{\mathrm{d}} \langle f^i_{t|t-1},f(z_t^{a^i}|\cdot)\rangle
}{
\lambda^\mathrm{fa}(z_t^{a^i})+P^{\mathrm{d}}\langle\lambda^u_{t|t-1},f(z_t^{a^i}|\cdot)\rangle
} \right\} \label{eq:JointEventProb2}
\end{align}
\else
\begin{align}
P_{t|t}&(a) \propto \prod_{i\in\{1,\dots,n_{t|t-1}\}|a^i=0} \left\{ 1- r^i_{t|t-1}P^{\mathrm{d}} \right\} \notag\\
& \times \prod_{i\in\{1,\dots,n_{t|t-1}\}|a^i>0} \left\{\frac{
r^i_{t|t-1}P^{\mathrm{d}} \langle f^i_{t|t-1},f(z_t^{a^i}|\cdot)\rangle
}{
\lambda^\mathrm{fa}(z_t^{a^i})+P^{\mathrm{d}}\langle\lambda^u_{t|t-1},f(z_t^{a^i}|\cdot)\rangle
} \right\} \label{eq:JointEventProb2}
\end{align}
\fi
which differs from \cite[p179]{ChaMor11} only in the inclusion of the term $P^{\mathrm{d}}\langle\lambda^u_{t|t-1},f(z_t^{a^i}|\cdot)\rangle$ in the denominator. Thus it is seen that JITS/JIPDA neglects the influence of unknown targets in the calculation of probabilities of association events.
From (\ref{eq:NewTargetNonExistWQ}), (\ref{eq:PoisUpdatePex}) and (\ref{eq:MargTrackFilter}), the posterior probability of existence of a track started on a measurement $z_t^j$ is ($i=n_{t|t-1}+j$, $j\in\{1,\dots,m_t\}$)
\begin{equation}\label{eq:InitialExistenceProbability}
r^i_{t|t} = \frac{P_{t|t}^i(2)P^{\mathrm{d}}\langle\lambda^u_{t|t-1},f(z_t^j|\cdot)\rangle}{\lambda^\mathrm{fa}(z_t^j) + P^{\mathrm{d}}\langle\lambda^u_{t|t-1},f(z_t^j|\cdot)\rangle} 
\end{equation}
This is exactly \cite[p325, (9.13)]{ChaMor11}, with $\rho_{\textrm{target}}(z)=P^{\mathrm{d}}\langle\lambda^u_{t|t-1},f(z|\cdot)\rangle$. Thus, track initiation in the variant of JITS/JIPDA modelling a birth intensity $\rho_{\textrm{target}}(z)$ can be seen to be equivalent to TOMB/P assuming a stationary intensity of unknown targets.\footnote{We also incorporate the kinematic distribution of targets into the initial update in (\ref{eq:PoisUpdateKin}); this distribution is assumed to be uniform in \cite{ChaMor11}.} The benefit of dynamically estimating this time-varying quantity was demonstrated in \cite{Wil12F2}.
Accordingly, we have shown that after two minor modifications and the approximation of (\ref{eq:MargTrackFilter}), JITS/JIPDA may be derived (and extended) using the RFS formalism. Similar algorithms have also been proposed within the RFS community; for example, \cite{MorCha03} updates existing tracks in a similar manner, but uses a two-point differencing approach to calculate initial states, weighting the new target hypotheses using a constant density of new targets (motivated as a truncated Poisson); and \cite{RisVo10} considers a bank of joint target-detection and tracking (JoTT) filters interacting through modified weights. 

The approximation made in the TOMB/P can be easily seen to preserve the first moment of the distribution. The first moment of the form (\ref{eq:LCMBIndepApprox}) can be written as:

\begin{align}
D_{t|t}(x) &= \sum_{a\in{\cal A}^{t|t}} P_{t|t}(a) \sum_{i\in{\cal T}_{t|t}}r_{t|t}^{i,a^i}f_{t|t}^{i,a^i}(x)  \\
& = \sum_{i\in{\cal T}_{t|t}} \left(\sum_{\tilde{a}\in{\cal A}^{t|t}|\tilde{a}^i=a^i} P_{t|t}(a)\right) r_{t|t}^{i,a^i}f_{t|t}^{i,a^i}(x)  \\
&= \sum_{i\in{\cal T}_{t|t}} \sum_{a^i\in{\cal H}^i_{t|t}} P^i_{t|t}(a^i) r_{t|t}^{i,a^i}f_{t|t}^{i,a^i}(x) 
\label{eq:TOMBFirstMoment}
\end{align}
which can be seen to be the first moment of the TOMB/P approximation (\ref{eq:MargTrackFilter}).

While our derivation has studied the unlabelled case, TOMB/P maintains tracks that are constructed from hypotheses which all begin with the same first detection of the target. Thus track continuity is implicitly maintained in the same way as in JPDA and related methods. This can be made explicit by incorporating a label element into the underlying state space, as recently proposed in \cite{VoVo13}; it can easily be shown that each track maintained by TOMB/P would retain a unique label. 

The only approximation made in the TOMB/P derivation was the approximation of the distribution $P_{t|t}(a)$ in (\ref{eq:MargTrackFilterWeights}). Practical implementations obviously require additional approximations in pruning tracks with small probability of existence, reducing the number of components in the mixture-conditioned distributions, and in calculating the marginal distributions $P_{t|t}^i(a^i)$. 

\subsection{Measurement oriented marginal MeMBer-Poisson filter}
\label{ss:MOMB}
{\noindent}The most striking difference between conventional filters and the MeMBer filter \cite{Mah07,VoVo09} is that the MeMBer collects all single target hypotheses updated with a particular measurement in the most recent scan into one Bernoulli component. This is the opposite of the TOMB/P (and most conventional trackers), which favours continuity of historical components over the constraints relating to the most recent measurement scan. An alternative MeMBer filter can be obtained by using a different marginalisation of $P_{t|t}(a)$ that follows this philosophy. We refer to the result as the measurement oriented marginal MeMBer/Poisson (MOMB/P) filter. For consistency, we continue to use the term \emph{track} to refer to the Bernoulli components although, unlike the TOMB/P, there is no correspondence of tracks between time steps.\footnote{As opposed to definition \ref{def:Track}, which defines a track as a collection of single-target hypotheses which all \emph{begin} with the same measurement, the Bernoulli components maintained by the MOMB/P consist of single-target hypotheses which all \emph{end} with the same measurement, i.e., all hypothesise the same final detection.}

The MOMB/P is based on an alternative parameterisation of association hypotheses.\footnote{Since we assumed that the prior was multi-Bernoulli, our hypotheses only consider the latest time step. For clarity of explanation, we initially omit missed detections, new targets and false alarms.} Rather than the parameterisation $a=(a^1,\dots,a^{n_{t|t}})$ where $a^i$ indexes the measurement associated with track $i$, we formulate hypotheses as $b=(b^1,\dots,b^{m_t})$ where $b^j$ indexes the \emph{track} associated with \emph{measurement} $j$, \textit{e.g.}\xspace, $b^j=i$ indicates that measurement $j$ is associated with track $i$. Under the assumption in (\ref{eq:MultiBernoulliAssumption}), $b$ is an equivalent parameterisation of the global association hypotheses. However, it gives rise to an alternative approximation of the distribution. Specifically, rather than seeking the best-fitting distribution that enforces independence in the association variables for different \emph{tracks} (\textit{i.e.}\xspace, (\ref{eq:TOMKLOpt})), we seek the best-fitting distribution that enforces independence in the association variables for different \emph{measurements}:
\[
P_{t|t}(b) \approx \prod_{j=1}^{m_t}q^{j*}(b^j)
\]
\begin{equation}\label{eq:MOMKLOpt}
\left[q^{j*}(b^j)\right]_j = \argmin_{q^j(b^j)}D\bigg(P_{t|t}(b)\bigg|\bigg|\prod_{j=1}^{m_t} q^j(b^j) \bigg)
\end{equation}
where, again following \cite[p277]{KolFri09}, $q^{j*}(b^j) =  P_{t|t}^j(b^j) = \sum_{\tilde{b}|\tilde{b}^j=b^j}P_{t|t}(\tilde{b})$ is the marginal probability that measurement $j$ is associated with track $i$.

\begin{figure}[tb]
\centering
\includegraphics[width=3.3in]{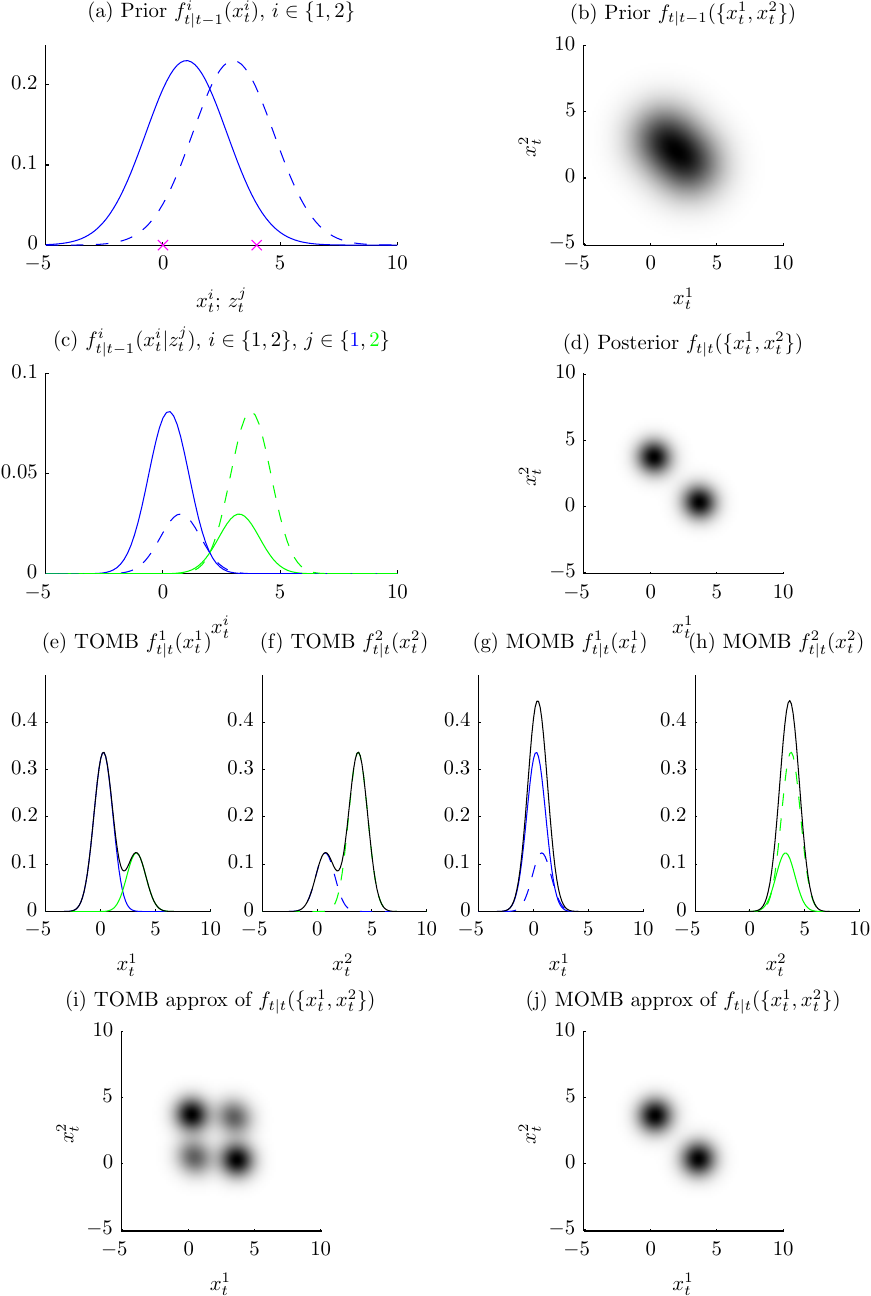}
\caption{Scenario involving two targets in one dimension. Prior tracks are shown in (a) (solid is $i=1$, dashed is $i=2$), along with measurements (``{\color{magenta}$\times$}''; left is $j=1$, right is $j=2$), The two-target joint prior distribution is shown in (b). Result of updating each component with each measurement is shown in (c) ({\color{blue}blue} is $j=1$, {\color{green}green} is $j=2$). Joint two-target posterior is shown in (d). Tracks carried forward by TOMB/P are shown in (e) and (f), and those carried forward by MOMB/P are shown in (g) and (h). Joint posterior approximations carried forward by TOMB/P and MOMB/P are shown in (i) and (j) respectively.}
\label{fig:MarginalisationDiagram}
\end{figure}

To illustrate the difference between the TOMB/P and MOMB/P, consider the case in Fig.~\ref{fig:MarginalisationDiagram}, involving two targets and two measurements, neglecting the possibilities of new targets or missed detections. In this simplified case, the multi-object RFS densities can be viewed as symmetric PDFs in the joint (two-target) state space. The prior distribution for each track and the measurement locations (marked as ``{\color{magenta}$\times$}'') are illustrated in (a). The corresponding two-target joint distribution is shown in (b); this is the symmetrisation of the distribution $f^1(x_t^1)f^2(x_t^2)$ (leaving off conditioning on prior measurements for simplicity), \textit{i.e.}\xspace,
\[
f^1(x_t^1)f^2(x_t^2) + f^2(x_t^1)f^1(x_t^2)
\]
The hypothesis-conditioned updated distributions for each prior marginal, updated with each measurement are shown in (c), weighted by the association likelihood $w^{i,j}=\int{f(z_t^j|x)f^i(x)\mathrm{d} x}$. The exact posterior for two targets is shown in (d); this is the symmetrised version of the distribution 
\[
w^{1,1}w^{2,2}f^1(x_t^1|z_t^1)f^2(x_t^2|z_t^2) + w^{1,2}w^{2,1}f^1(x_t^1|z_t^2)f^2(x_t^2|z_t^1)
\]
The first term in the sum above is the product of the two larger peaks (blue solid, green dashed), while the second is the product of the smaller peaks (blue dashed, green solid). Because $f^1(\cdot|z_t^1)$ is similar to $f^2(\cdot|z_t^1)$ (and $f^1(\cdot|z_t^2)$ is similar to $f^2(\cdot|z_t^2)$), the second term is similar to the first term, but with target identities switched. After symmetrising, we arrive at the joint in (d).

The posterior tracks retained  by the TOMB/P (and JIPDA, \textit{etc}\xspace) are shown in (e) and (f); this yields the approximation of the posterior in (i), which is a symmetrised version of the distribution
\ifCLASSOPTIONdraftcls
\[
\big[p^{1,1}f^1(x_t^1|z_t^1) + p^{1,2}f^1(x_t^1|z_t^2)\big]\cdot
\big[p^{2,1}f^2(x_t^2|z_t^1) + p^{2,2}f^2(x_t^2|z_t^2)\big]
\]
\else
\begin{multline*}
\big[p^{1,1}f^1(x_t^1|z_t^1) + p^{1,2}f^1(x_t^1|z_t^2)\big] \\
\times \big[p^{2,1}f^2(x_t^2|z_t^1) + p^{2,2}f^2(x_t^2|z_t^2)\big]
\end{multline*}
\fi
where $p^{i,j}=P_{t|t}^i(a^i=(t,j))=P_{t|t}^j(b^j=i)$ is the marginal probability that measurement $j$ is associated with track $i$. The previous section showed this to be the best approximation of the association variables in which the associations for different \emph{tracks} are forced to be independent. 

In (\ref{eq:MeMBer}), we will see that the posterior tracks retained by the measurement-oriented marginalisation in (\ref{eq:MOMKLOpt}) are those shown in (g) and (h). The corresponding approximation of the joint distribution is a symmetrisation of
\ifCLASSOPTIONdraftcls
\[
\big[p^{1,1}f^1(x_t^1|z_t^1) + p^{2,1}f^2(x_t^1|z_t^1)\big]\cdot
\big[p^{1,2}f^1(x_t^2|z_t^2) + p^{2,2}f^2(x_t^2|z_t^2)\big]
\]
\else
\begin{multline*}
\big[p^{1,1}f^1(x_t^1|z_t^1) + p^{2,1}f^2(x_t^1|z_t^1)\big]\\
\times\big[p^{1,2}f^1(x_t^2|z_t^2) + p^{2,2}f^2(x_t^2|z_t^2)\big]
\end{multline*}
\fi
as illustrated in (j). In this instance, the MOMB/P approximation clearly provides a better representation of the true distribution.

Before proceeding, we expand our notation to admit new targets, false alarms and missed detections. Since we have assumed that the prior distribution was multi-Bernoulli (\ref{eq:MultiBernoulliAssumption}), posterior hypotheses will be either ${\cal M}^t(i,a^i)=\emptyset$ (the target was missed) or ${\cal M}^t(i,a^i)=\{(t,j)\}$ (measurement $z_t^j$ corresponds to target $i$). This set of hypotheses can be characterised equivalently via a set of hypotheses $b=(b^1,\dots,b^{m_t+n_{t|t-1}})$ where for $j\in\{1,\dots,m_t\}$, $b^j=0$ if measurement $j$ is hypothesised to not correspond to a previously detected target, and $b^j=i$ if measurement $j$ corresponds to pre-existing target $i$, and for $j=m_t+i$, $i\in\{1,\dots,n_{t|t-1}\}$, $b^{m_t+i}\in\{0,i\}$, where $b^{m_t+i}=i$ if track $i$ is missed (and $0$ otherwise). The global hypotheses $a$ and $b$ are alternative parameterisations of the same event. Thus (\ref{eq:LCMBIndepApprox}) can be written equivalently as
\begin{align}
f^\mathrm{mbm}_{t|t}(X) &= \sum_{\alpha\in{\cal P}_{n_{t|t}}^{|X|},b\in{\cal B}_t} P_{t|t}(b) \label{eq:MOMFullDistribution} \prod_{j=1}^{m_t+n_{t|t-1}}f_{t|t}^{\beta(b^j,j)}(X_{\alpha(j)}) \\
P_{t|t}(b) &\propto \prod_{j=1}^{m_t+n_{t|t-1}}w_{t|t}^{\beta(b^j,j)}, \quad b\in{\cal B}
\end{align}
where, in analogy with (\ref{eq:AssocAlphabet}),
\ifCLASSOPTIONdraftcls
\begin{multline}\label{eq:MOAssocAlphabet}
{\cal B}_t = \bigg\{(b^1,\dots,b^{m_t+n_{t|t-1}}) \bigg| 
  b^j\in\{0,\dots,n_{t|t-1}\},j\in\{1,\dots,m_t\}, 
  b^{m_t+i}\in\{0,i\},i\in\{1,\dots,n_{t|t-1}\}, \\
  \{1,\dots,n_{t|t-1}\}\subseteq{\textstyle\bigcup_{j=1}^{m_t+n_{t|t-1}}\{b^j\}}, 
  b^j\neq b^{j'} \forall j\neq j' \mbox{s.t. } b^j\neq 0
\bigg\}
\end{multline}
\else
\begin{multline}\label{eq:MOAssocAlphabet}
{\cal B}_t = \bigg\{(b^1,\dots,b^{m_t+n_{t|t-1}}) \bigg| \\
  b^j\in\{0,\dots,n_{t|t-1}\},j\in\{1,\dots,m_t\}, \\
  b^{m_t+i}\in\{0,i\},i\in\{1,\dots,n_{t|t-1}\}, \\
  \{1,\dots,n_{t|t-1}\}\subseteq{\textstyle\bigcup_{j=1}^{m_t+n_{t|t-1}}\{b^j\}}, \\
  b^j\neq b^{j'} \forall j\neq j' \mbox{s.t. } b^j\neq 0
\bigg\}
\end{multline}
\fi
and $\beta(b^j,j)$ maps to the hypothesis and track corresponding to $b^j$:
\begin{itemize}
\item if $b^j=i\in\{1,\dots,n_{t|t-1}\}$ for $j\in\{1,\dots,m_t\}$ then $\beta(b^j,j)$ maps to the hypothesis in (\ref{eq:DetUpdateMeasSet})-(\ref{eq:DetUpdateKin}) updating track $i$ with measurement $j$
\item if $b^j=0$ for $j\in\{1,\dots,m_t\}$ then $\beta(b^j,j)$ maps to the new target hypothesis in (\ref{eq:PoisUpdateMeasSet})-(\ref{eq:PoisUpdateKin})
\item if $b^j=i$ for $j=m_t+i$ then $\beta(b^j,j)$ maps to the missed detection hypothesis in (\ref{eq:MissUpdateMeasSet})-(\ref{eq:MissUpdateKin})
\item if $b^j=0$ for $j=m_t+i$ then $\beta(b^j,j)$ maps to the ``non-existence'' hypothesis, \textit{i.e.}\xspace, (\ref{eq:NewTargetNonExistWQ})
\end{itemize}
Subsequently, we obtain an alternative approximation based on the representation of $P_{t|t}(b)$ via the product of its marginals:
\begin{equation}\label{eq:MOApprox}
P_{t|t}(b) \approx \prod_{j=1}^{m_t+n_{t-1}}P_{t|t}^j(b^j), \quad P_{t|t}^j(b^j) = \sum_{\tilde{b}\in{\cal B}_t|\tilde{b}^j=b^j}P_{t|t}(\tilde{b})
\end{equation}
Again, with no further approximation, the joint distribution is multi-Bernoulli:
\begin{equation}
\begin{split}
f^\mathrm{mbm}_{t|t}(X) &\approx \sum_{\alpha\in{\cal P}_{n_{t|t}}^{|X|}}\prod_{j=1}^{m_t+n_{t-1}}
f_{t|t}^j(X_{\alpha(j)}) \\
f_{t|t}^j(X) &= \sum_{b^j}P_{t|t}^j(b^j)
f_{t|t}^{\beta(b^j,j)}(X)
\end{split}
\label{eq:MeMBer}
\end{equation}
We have thus obtained an alternative form of update for a filter that is structurally similar to the MeMBer \cite{Mah07,VoVo09}, i.e., it collects all single-target hypotheses updated with a given measurement into a single posterior Bernoulli component (track). The difference between the MeMBer and MOMB/P is in the weights applied to each hypothesis: the former arises from approximations made to the p.g.fl \cite[p668]{Mah07}, while the latter interprets the terms in the sum over $b\in{\cal B}_t$ in (\ref{eq:MOMFullDistribution}) as a distribution over association events. The only approximation made in the derivation is the approximation of the joint association distribution $P_{t|t}(b)$ as the product of its marginals.\footnote{Again, practical implementations will require additional approximations in pruning unlikely tracks, reducing the number of components in the mixture-conditioned distributions, and in calculating the marginal distributions $P_{t|t}^j(b^j)$ themselves.}

\begin{figure}[tb]
\centering
\includegraphics{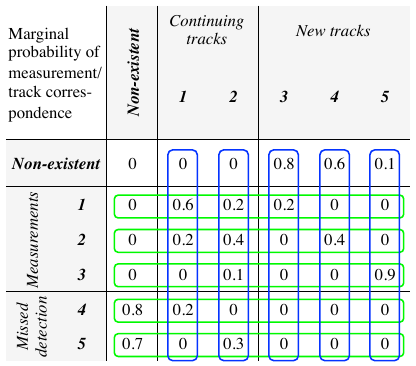} %[width=3.3in]
\caption{Example of marginal distribution of track-measurement association, and Bernoulli components formed by {\color{blue}TOMB/P} (collecting columns) and {\color{green}MOMB/P} (collecting rows). Each cell of the table contains the marginal probability that a given measurement is associated with a given track (which sums over the same global association hypotheses as the probability that the track is associated with the measurement). In the example, there are two continuing tracks and three new measurements; this results in five posterior tracks (\textit{i.e.}\xspace, Bernoulli components). Using the TOMB/P, these correspond to the two existing tracks, and a new track for each measurement. Using the MOMB/P, a posterior track exists for each measurement, collecting events from all prior tracks that use the measurement. The remaining posterior tracks hypothesise missed detection of the corresponding prior track (and non-existence otherwise).}
\label{fig:MarginalSlicing}
\end{figure}

The difference between the tracks formed by the TOMB/P and MOMB/P is illustrated in Fig.~\ref{fig:MarginalSlicing}. In the TOMB/P, there is a track in the posterior corresponding to each prior track collecting all hypotheses updating that track, and a track in the posterior for each measurement considering the possibility of the measurement being a new target. In the MOMB/P, there is a track for each measurement collecting all hypotheses \emph{updated by} the measurement (collecting hypotheses from all prior tracks), and a track for each prior track containing only the ``missed detection'' hypothesis.

Detailed pseudocode for both TOMB/P and MOMB/P can be found in Appendix \ref{app:Pseudocode}, and a simplified Matlab implementation can be found in the ancillary files to \cite{Wil14c}. While there is no inherent difference in the complexity of the two methods, the run times in the experiments differ slightly due to the different numbers of tracks and hypotheses that are maintained after pruning.

\subsection{Approximating marginal association distributions}
\label{ss:BP}
{\noindent}The greatest barrier to practical application of the TOMB and MOMB algorithms is the calculation of the marginal probabilities of association $P_{t|t}^i(a^i)$ and $P_{t|t}^j(b^j)$, which is closely related to the \#P-complete calculation of a matrix permanent. The present work was motivated by the emergence of a new approximation to this calculation that is both accurate and tractable \cite{WilLau12}. The approximation is based on loopy belief propagation (LBP), and the particular model used has been studied recently in machine learning, physics and information theory.\footnote{See \cite{WilLau12,Von13} for references.} While LBP was discovered as a heuristic extension of the standard forwards-backwards algorithm for Markov chains to probabilistic graphs with loops, it has since been understood as a variational (\textit{i.e.}\xspace, optimisation-based) inference method with roots in statistical physics \cite{YedFre05}. Although there are few cases in which the performance of LBP is guaranteed, its practical performance in many problems is remarkable; for example, turbo decoding has been shown to be an instance of LBP \cite{McEMac98}. In general graphical models LBP is not guaranteed to converge; however, it was proven simultaneously in the conference papers preceding \cite{WilLau12,Von13} that LBP is guaranteed converge to a unique solution in the particular formulation of the data association problem that we utilise. The underlying optimisation problem was shown to be convex in \cite{Von13}. Details of the algorithm, the accuracy of the marginal estimates that it yields and the number of iterations required for convergence are provided in \cite{WilLau12}; detailed pseudocode for the algorithm and its employment in the TOMB/P and MOMB/P can be found in Appendix \ref{app:Pseudocode}, and a simplified Matlab implementation can be found in the ancillary files to \cite{Wil14c}. The complexity of the method per iteration can easily be seen to be $O(n_{t|t-1}m_t)$. As examined in \cite{WilLau12}, the number of iterations required depends on the problem parameters. The computation time associated with the method in our experiments is examined in Section~\ref{ss:ExpComputation}.

Note that the filter derivation above is not tied to the LBP approximation; other methods such as Maskell's exact EHM \cite{HorMas06} or MCMCDA \cite{OhRus09} may also be used to calculate the marginal association probabilities.

\section{Relationship to other work}
\label{sec:Relationship}
The relationships of the proposed algorithms to the JITS/JIPDA and MeMBer filters were discussed in Sections \ref{ss:TOMB} and \ref{ss:MOMB}. As discussed in Section~\ref{sec:Derivation}, the structure of the MBM component of the distribution is similar to that of the TOMHT (in its use of a hypothesis tree for each target, implicitly representing global association hypotheses). At each time, MHT methods seek the most probable global association hypothesis, which corresponds to the term in the sum (\ref{eq:PreexistingTargets}) with the largest weight.\footnote{As discussed in Section~\ref{sec:Derivation}, hypotheses in our derivation are multi-Bernoulli, and thus they imply a distribution over cardinality; in contrast, MHT hypotheses uniquely determine cardinality. This difference permits us to represent many MHT hypotheses (\textit{e.g.}\xspace, those proposing missed detection or death of the same targets) via a single multi-Bernoulli hypothesis.} The derivation in Section~\ref{sec:Derivation} retains separability of the weights (\ref{eq:GlobalHypWeight}) so that efficient methods such as Lagrangian relaxation could be applied to obtain a RFS-based TOMHT; these were applied in \cite{Wil11b,LauWil11}. 

It has been suggested that the explicit modelling of associations in MHT may induce a bias into the solution (\textit{e.g.}\xspace, \cite[p340]{Mah07}). While the derivation in Section~\ref{sec:Derivation} has shown that association naturally arises in RFS filters, it appears as a total probability expansion (\ref{eq:PreexistingTargets}), marginalising over the unobserved hypotheses. Thus the derivation does not validate approaches such as MHT that replace the sum over $a\in{\cal A}_{t|t}$ in (\ref{eq:PreexistingTargets}) with a maximisation.

The derivation is related to \cite{VoVo13}, which shows that the labelled case can be handled within the unlabelled framework by incorporating a label element in to the underlying state space. The formulation in \cite{VoVo13} is constructed to ensure that uniqueness of label is maintained; this uniqueness is then exploited in the derivation. On a practical level, both methods effectively involve a total probability expansion over an intractable number of terms (\textit{i.e.}\xspace, global association hypotheses). A significant contribution of the derivation in Section~\ref{sec:Derivation} is to show that global hypotheses in the full Bayes RFS filter can be represented in a manner similar to TOMHT, \textit{i.e.}\xspace, through a collection of single-target trees (a track oriented approach), enabling use of methods which exploit separability of weights (e.g., the proposed TOMB and MOMB algorithms, and the subsequent extension in \cite{Wil14}). In contrast, the filter in \cite{VoVo13} operates by carrying a fixed number ($N$) of terms, each of which corresponds to a global association hypothesis. The update step then applies Murty's algorithm to find the $N$ terms with the highest weight, truncating the sum in a manner similar to the MHT implementation in \cite{Cox96}. However, the potential bias in MHT is averted by finding the MAP cardinality estimate, and then utilising the highest weight component with that cardinality (conceptually similar to the MaM estimator in \cite[p497]{Mah07}). The efficiency of the approach is improved in \cite{VoVo14}, using a $k$-th shortest path algorithm to avoid enumeration of all global hypotheses upon prediction.

The concept of maintaining a representation of targets that have never been detected is uncommon but not new; \textit{e.g.}\xspace, see \cite{MorCho86}. While this aspect of the filter has not been a focus in this paper, it yields significant benefits including in the speed of initiation during the initial transient when the tracker commences and in environments with non-stationary sensor coverage \cite{Wil12F2}. Furthermore, it is shown in \cite{Wil12F2} that tracks with a low probability of existence can be efficiently represented via the Poisson component, greatly reducing the complexity associated with initiating tracks in high false alarm environments.

\section{Results}
\label{sec:Results}
We demonstrate the performance of the proposed approaches in challenging scenarios involving $n\in\{6,10,20\}$ targets which are in close proximity at the mid-point of the simulation, achieved by initialising at the mid-point and running forward and backward dynamics. We consider two cases for the mid-point initialisation (\textit{i.e.}\xspace, $t=100$):
\begin{align*}
\mbox{Case 1: } & x_{100} \sim \mathcal{N}\{0,10^{-6}\times\mathbf{I}_{4\times 4}\} \\
\mbox{Case 2: } & x_{100} \sim \mathcal{N}\{0,0.25\times\mathbf{I}_{4\times 4}\} 
\end{align*}
\begin{figure}[tb]
\centering
\includegraphics[width=3.3in]{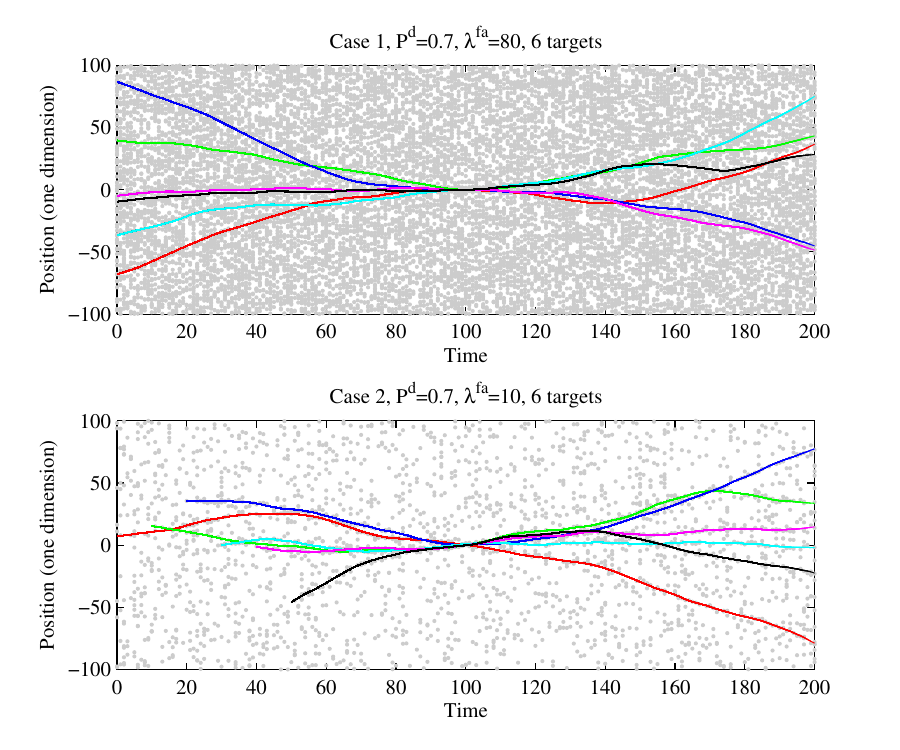}
\vspace{-6pt}
\caption{One dimension of a single Monte Carlo run of the scenario cases 1 and 2. Target trajectories shown in colours, and measurements shown in gray. Both use $P^{\mathrm{d}}=0.7$; top and bottom have the expected number of false alarms set to $80$ and $10$ respectively.}
\label{fig:Scenario}
\vspace{-12pt}
\end{figure}%
where the target state is position and velocity in two dimensions. Snapshots of one dimension of both cases are shown in Fig.~\ref{fig:Scenario}. It is well-known that JPDA and related algorithms suffer from \emph{coalescence}, \textit{i.e.}\xspace, after targets come in close proximity, mean estimates remain on the mid-point of the targets for a significant duration (\textit{e.g.}\xspace, \cite{BloBlo00}). In case 1, targets are completely indistinguishable at the mid-point, hence coalescence-like effects are at their worst. Coalescence-like effects also occur in case 2, but to a lesser extent (mainly due to the discernible difference in velocity). In case 1, targets all exist throughout the simulation (tracks are not pre-initialised). In case 2, the  targets are born at times $\{0,10,\dots,10 (n-1)\}$ (any targets not existing prior to time $t=100$ are born at that time; consequently, for case 2 with $n=20$, ten targets are born at time $t=100$). Targets follow a linear-Gaussian model with nominally constant velocity, $x_t = \mathbf{F} x_{t-1} + w_t$, where $w_t\sim\mathcal{N}\{0,\mathbf{Q}\}$, 
\[
\mathbf{F} = \left[\begin{array}{cc}
1 & T \\
0 & 1
\end{array}\right] \otimes \mathbf{I}_{2\times 2}, \quad
\mathbf{Q} = q \left[\begin{array}{cc}
T^3/3 & T^2/2 \\
T^2/2 & T
\end{array}\right] \otimes \mathbf{I}_{2\times 2}
\]
and $q = 0.01$, $T=1$. 
Target-originated measurements provide position corrupted by Gaussian noise, \textit{i.e.}\xspace, $z_t = \mathbf{H} x_t + v_t$, where $\mathbf{H} = [\;1 \;\; 0\;] \otimes \mathbf{I}_{2\times 2}$, and $v_t\sim\mathcal{N}\{0,\mathbf{I}_{2\times 2}\}$. The initial unknown target intensity for the proposed methods is assumed to be $\lambda^u_{0|0}(x)=10\mathcal{N}\{x;0,\mathbf{P}\}$, where $\mathbf{P}=\mathrm{diag}[100^2,\; 1,\; 100^2,\; 1]$ covers the position and velocity region of interest. The birth intensity uses the same covariance, $\lambda^\mathrm{b}(x)=0.05\mathcal{N}\{x;0,\mathbf{P}\}$. Cases are considered with the average number of false alarms per scan as $\lambda^\mathrm{fa}\in\{10,20,40,60,80\}$, and their distribution is uniform on $[-100,100]^2$. All filters assume $P^{\mathrm{s}}=0.999$. We consider cases with $P^{\mathrm{d}}\in\{0.3,0.5,0.7,0.98\}$, representing a range of SNR values.  

The scenarios examined are exceptionally challenging due to the large number of targets in close proximity. While others have considered larger numbers of targets, in most cases these positioned uniformly in space and rarely come into close contact. Cases such as this can be effectively decoupled into series of single target tracking problems. In the present study, up to 20 targets have effectively the same position and velocity at the mid-point in time, and the dependency between targets is inescapable. 

We compare the proposed methods to CPHD \cite{Mah07,VoVo07}, CB-MeMBer \cite{VoVo09}, JITS and linear multi-target ITS (LMITS) \cite{MusEva09,ChaMor11} (a variant of JITS which replaces the marginal association probabilities with a heuristic approximation). The CB-MeMBer approximates the initial distribution by 30 identical Bernoulli components with $r=1/3$, and uses a single birth component with $r=0.05$ at each time. JITS and LMITS truncate the unknown target intensity to use $\rho_{\textrm{target}}(z)=P^{\mathrm{d}}\langle\lambda^\mathrm{b},f(z|\cdot)\rangle$. The implementation of JITS initiates a new track on every measurement (via (\ref{eq:InitialExistenceProbability})), but only updates new tracks with measurements that are not in the gate of any hypothesis of an existing track; this was necessary for tractability. Exact marginal association probabilities were calculated using the junction tree algorithm described in \cite{WilLau12}, which in turn uses the library for discrete approximate inference \cite{libDAI}. The exact calculation is not possible for the simulations with $n>6$ targets, or with $\lambda^\mathrm{fa}>10$. Since the weights provided by LMITS do not provide a probability that a measurement is not used by any other track, the LMITS implementation initiates tracks only on measurements that are not within the gate of any hypothesis.  
Where applicable, all tracks with $r^i_{t|t}\geq 10^{-4}$ are maintained (the threshold needed to be increased to $10^{-3}$ for JITS to be able to successfully execute). All methods utilise Gaussian mixture implementations, with hypothesis management performed using $n$-scan merging \cite{SinSea74}, with $n=2$.\footnote{Histories are only considered to be identical if the last two measurement updates are the same---\textit{i.e.}\xspace, missed detection events are not considered to be an element in the history. A maximum of 1000 single-target hypotheses are retained (in total), and hypotheses with $w^{i,a^i}_{t|t}r^{i,a^i}_{t|t}\leq 10^{-4}$ are deleted.} All methods tested share the same filtering and hypothesis management code, so run times are somewhat comparable. All code was written in MATLAB except for the C++ junction tree routine used for JITS \cite{libDAI}. Clustering is used only in the JITS marginal probability calculation. The TOMB/P, JITS and LMITS implementations output estimates\footnote{The estimate is the mean of the Gaussian component with the highest weight.\label{fn:Estimate}} for all tracks with existence probability $\geq 0.8$, while the remaining methods find the mode of the cardinality distribution (excluding the Poisson component for the MOMB/P), and output estimates\footref{fn:Estimate} for the corresponding number of tracks with the highest probability of existence (or, in the case of CPHD, the Gaussian components with highest weight).

\subsection{Tracking performance}
\label{ss:ExpPerformance}
The results are shown in Fig.~\ref{fig:Results}; the figure shows the mean optimal sub-pattern assignment (OSPA)\footnote{OSPA provides a single metric for measuring distance between two finite sets, thus summarising errors in both position and cardinality in a single value.} \cite{SchVo08} with $p=1$ and $c=20$ (calculated over both position and velocity) as a function of time, averaged over $200$ Monte Carlo trials. The following observations are made:
\begin{figure*}[p]
\centering
\ifCLASSOPTIONdraftcls
\includegraphics[width=6.5in]{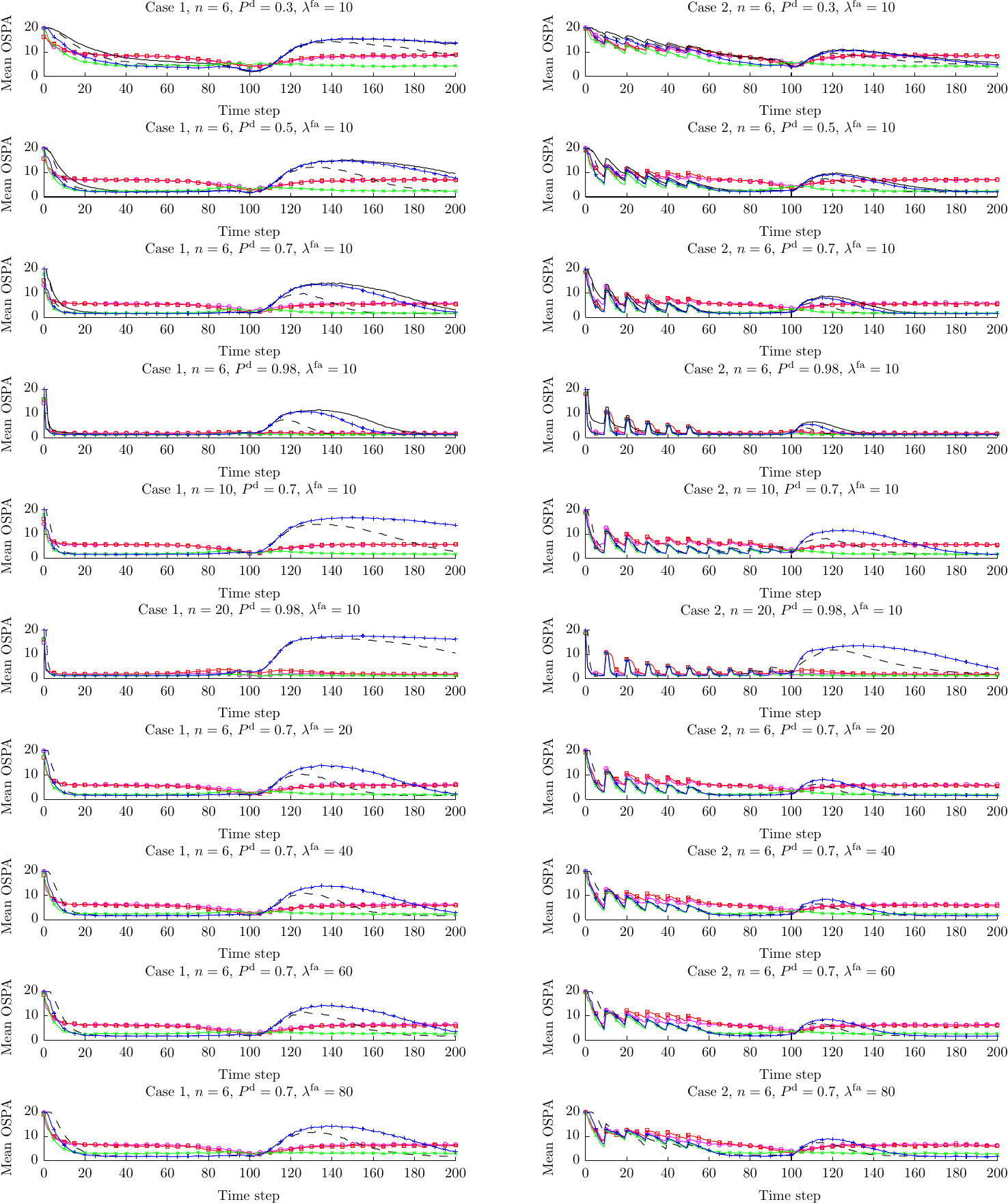}
\else
\includegraphics[width=7in]{figure5.pdf}
\fi
\vspace{-6pt}
\caption{Performance of tracking methods on scenarios involving $n\in\{6,10,20\}$ targets. Targets are in close proximity at time $100$ (to the point of being indistinguishable in case 1). Algorithms are encoded by colour and symbol as TOMB/P {\color{blue}$+$}, MOMB/P {\color{green}$\times$}, CB-MeMBer {\color[rgb]{1,0,1}$\circ$}, CPHD {\color{red}$\Box$}, JITS (solid black) and LMITS (dashed black). JITS is excluded from cases with $n>6$ and $\lambda^\mathrm{fa}>10$ as it could not complete these cases.}
\label{fig:Results}
\vspace{-6pt}
\end{figure*}
\afterpage{\clearpage}
\begin{itemize}
\item The effect of coalescence is clearly visible from $t=100$ to the end of the simulation for JITS, TOMB/P and LMITS. The cause of this difficulty is illustrated in figure \ref{fig:MarginalisationDiagram}(i): after targets have been closely spaced, the posterior marginal distribution for each track may contain a mode representing each target. Thus the product of these marginal distributions contains ``alias'' peaks with multiple targets in the same region, and in the output of the tracker, multiple estimates may be placed on the same target. Interestingly, LMITS is affected less than TOMB/P, which is affected less than JITS. The difference in each of these is the approximation of the marginal association probabilities used---the approximation used by LMITS is shown in \cite{WilLau12} to be inferior to the LBP-based method used for the TOMB/P, whereas the JITS uses the exact values. Counter-intuitively, the poorer approximations perform better. This suggests that the approximations tend to be less equivocal between the competing hypotheses than the exact probabilities, e.g., giving higher probability to the most likely hypothesis, and lower probability to alternative hypotheses, thus aiding earlier resolution of coalescence. A similar difference between JIPDA and LMIPDA was observed in \cite{MusLaS08}. Various methods exist for modifying association weights to resolve coalescence (\textit{e.g.}\xspace, \cite{BloBlo00,SveSve11}), but all suffer from similar exponential complexity to the exact marginal association calculation. An extension of the current paper which provides a tractable approximation of the multi-Bernoulli distribution that minimises the RFS KL divergence from the full distribution is available in \cite{Wil14}.

\item MOMB/P (as well as CPHD and CB-MeMBer) exhibits good robustness to coalescence effects. An example of this was illustrated in figure \ref{fig:MarginalisationDiagram}(j): when coalescence effects are present, the MOMB approximation of the association distribution (\ref{eq:MOApprox}) yields a multi-Bernoulli RFS density which tends to be closer to the full RFS density. This is due to the fact that all hypotheses in the Bernoulli component $j\in\{1,\dots,m_t\}$ have been updated with measurement $z_t^j$, hence the components tend to be spatially concentrated. Consequently, the product of the marginal distributions will tend not to contain ``alias'' peaks, as shown in figure \ref{fig:MarginalisationDiagram}(j).

\item CPHD, CB-MeMBer and MOMB/P also react more quickly to cardinality changes due to the use of the cardinality mode estimator rather than component-by-component thresholding. Where coalescence is not dominant, all of these methods exhibit reduced performance in comparison to the various track oriented approaches. The reduction in performance is larger in lower $P^{\mathrm{d}}$ cases. The MOMB/P outperforms CPHD and CB-MeMBer throughout, especially in the low $P^{\mathrm{d}}$ cases. The primary error in CPHD and CB-MeMBer in these cases is the so-called ``spooky effect'' \cite{FraSch09}, \textit{i.e.}\xspace, in cases in which one or more targets are not detected, their positions are estimated to be close to other targets that are detected, regardless of target spacing. Thus, as described in \cite{FraSch09}, the global cardinality is estimated accurately but the local cardinality is inaccurate. Neither TOMB/P nor MOMB/P exhibit the same difficulty. One possible explanation of this is that, while CB-MeMBer retains a multi-Bernoulli distribution, the cardinality correction is based on the overall intensity, hence, like CPHD, some information on local cardinality is lost. The approximation used in the proposed methods does not utilise the global intensity in the same way.

\item With $P^{\mathrm{d}}=0.3$, TOMB/P, MOMB/P and CB-MeMBer all maintain around $200$ tracks on average, whereas JITS and LMITS maintain $20$--$30$. This is the consequence of the track initiation logic necessitated by JITS and LMITS, and the higher track deletion threshold for JITS (which was necessary to successfully execute the algorithm); it directly leads to the degradation in performance of these methods compared to the others with this value of $P^{\mathrm{d}}$. Thus the apparent lower complexity of JITS in Fig.~\ref{fig:TimeResults} is the result of its solving a much smaller problem, which leads to its lower performance, due to slow initiation. Scenarios with $n\in\{10,20\}$ targets or $\lambda^\mathrm{fa}>10$ could not be executed at all for JITS due to the large memory and computation requirements. The extension of the current paper in \cite{Wil12F2} demonstrates how the PPP component can be used to represent tracks with a low probability of existence to avoid maintaining such a large number of tracks.
\end{itemize}

\subsection{Computational complexity}
\label{ss:ExpComputation}
\begin{figure*}[!t]
\centering
\includegraphics[width=7in]{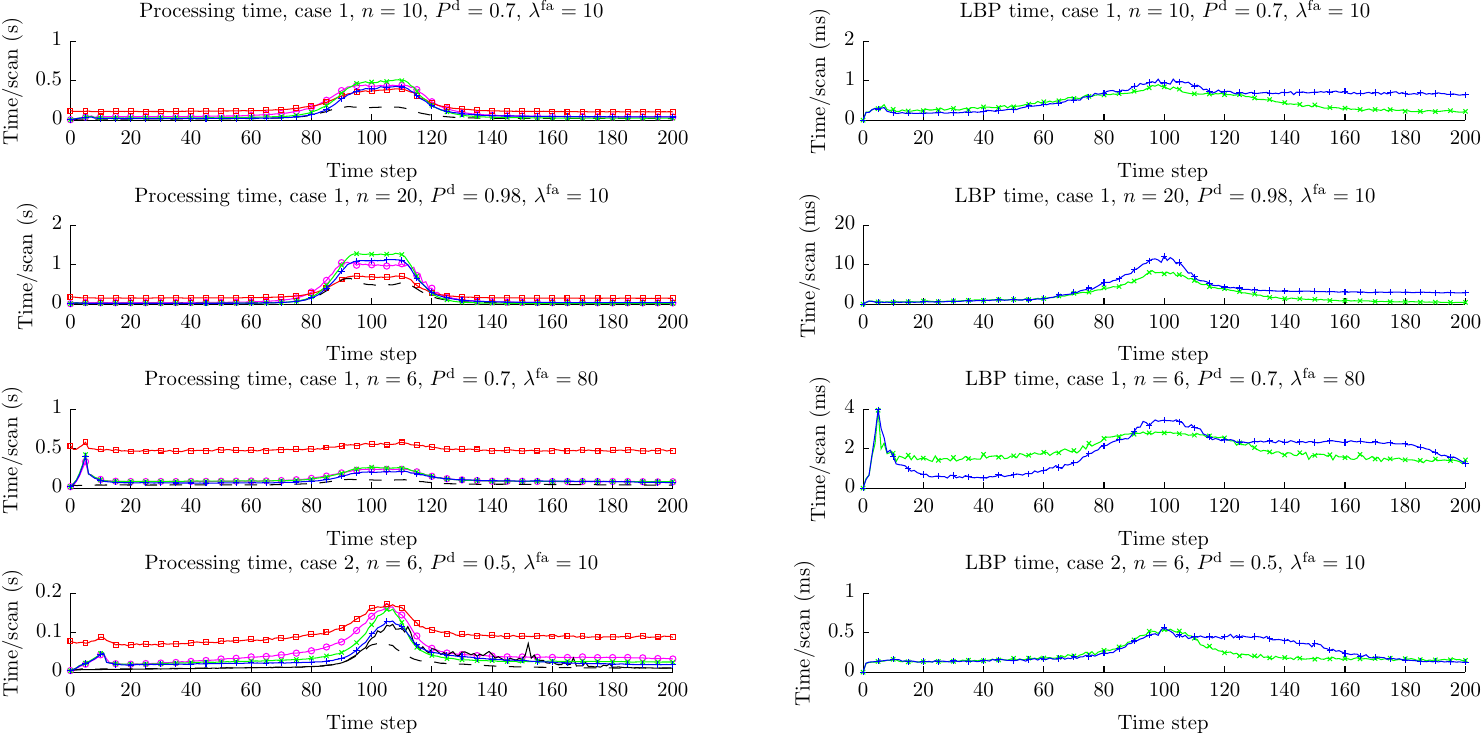}
\vspace{-6pt}
\caption{Average computation time per time step (left) for various methods, and average time for completion of LBP algorithm (right). Algorithms are encoded by colour and symbol as TOMB/P {\color{blue}$+$}, MOMB/P {\color{green}$\times$}, CB-MeMBer {\color[rgb]{1,0,1}$\circ$}, CPHD {\color{red}$\Box$}, JITS (solid black) and LMITS (dashed black). JITS is excluded from cases with $n>6$ and $\lambda^\mathrm{fa}>10$ as it could not complete these cases.}
\label{fig:TimeResults}
\vspace{-6pt}
\end{figure*}
The computation time for the various algorithms on a subset of scenarios is examined in Fig.~\ref{fig:TimeResults}. The left column show the total processing time for each scan (\textit{i.e.}\xspace, each time step of the simulation) using each algorithm. While the computation times are quite long, the vast majority of the time is spent in calculation of a large number of Gaussian mixture hypotheses. For example, in the scenario with $n=20$ targets, the maximum number of $1000$ single-target hypotheses will be maintained around the mid-point in time. Each of these hypotheses needs to be updated with each of the target-originated measurements in the next time step (since the targets are closely spaced), yielding around $20,\!000$ hypotheses which are subsequently simplified back to $1000$. The times spent in this process differ slightly for the various methods, as the differences in hypothesis weights yield variations in the number of hypotheses and tracks that need to be maintained. More sophisticated hypothesis reduction methods have been successfully applied, but are not presented here as they cannot easily be applied to the comparison algorithms (CPHD and CB-MeMBer).

The results with $\lambda^\mathrm{fa}=80$ show that the complexity of CPHD is strongly affected by the total number of measurements, whereas the other methods are more impacted by the number of targets in close proximity. Through implementation methods such as gating and clustering (which are widely used in practical systems, \textit{e.g.}\xspace, \cite{BlaPop99}), complexity scales very well to very large problems involving well-spaced targets (or small groups of well-spaced targets) since the sub-problems in each cluster are independent.

The only operation which is performed by TOMB/P and MOMB/P but not the other methods is the LBP approximation of marginal association probabilities. The total time consumed in the iterative calculation for each scan of the simulations is shown in the right column of Fig.~\ref{fig:TimeResults} (note that time is in milli-seconds, and clustering is not applied). Even in the most challenging case ($n=20$), average computation time peaks at $15$\ ms; the values in other cases are significantly smaller. As discussed previously, the algorithms as stated maintain a large number of tracks with low probability of existence. The extension of the current paper in \cite{Wil12F2} demonstrates that these can be efficiently represented using the PPP component, resulting in a significant computational saving.

\section{Conclusion and Extensions}
\label{sec:Conclusion}
In this paper, we have presented a derivation of a conjugate prior for a full Bayes RFS filter under common assumptions, and shown that the resulting data structure is similar to that utilised in TOMHT. Subsequently, the TOMB and MOMB methods were derived based on approximations of the marginal association distribution, which can be efficiently approximated using the LBP algorithm. Experimental scenarios demonstrated promising performance on a challenging problem. 

By comparing the resulting TOMB/P algorithm with JITS/JIPDA, we showed how variants of these popular methods can be derived using the RFS framework. The newly-proposed MOMB/P algorithm follows the philosophy of the MeMBer alongside the LBP data association approximation to provide a solution that is robust to coalescence, and considerably improves upon the performance of CPHD and CB-MeMBer, especially with lower $P^{\mathrm{d}}$ values. 

To date, this work has been extended in three directions:
\begin{itemize}
\item In \cite{Wil12F2}, we demonstrate the significant computational benefit of representing Bernoulli components with a small probability of existence through the PPP component of the distribution.

\item In \cite{Wil14}, we derive a variational method for (approximately) finding the multi-Bernoulli distribution with the smallest KL divergence from the full RFS distribution (\ref{eq:PreexistingTargets}) (rather than minimising the KL divergence of the discrete association distribution, as in (\ref{eq:TOMKLOpt}) and (\ref{eq:MOMKLOpt})). The result combines the superior performance of TOMB/P when targets are well-separated with the robustness against coalescence of MOMB/P.

\item In \cite{BreKal14}, the derivation in Section~\ref{sec:Derivation} is extended to single-cluster processes, and the resulting method is applied to simultaneous localisation and mapping.
\end{itemize}

\appendices
\section{Proof of prediction step}
\label{app:PredictionProof}
\begin{IEEEproof}[Proof of Theorem~\ref{th:Prediction}]
The proof follows these steps:
\begin{enumerate}
\item Substitute the p.g.fl form of the density (\ref{eq:FullDensity}) into the Bayes RFS prediction formula (\ref{eq:PGFLPrediction})
\item Observe that the result is the product of two p.g.fls 
\item Show that the first component is a PPP, show that the intensity of the PPP is calculated using a standard PHD prediction step \cite{Mah03,VoMa06}
\item Show that the second component is a MBM, and that the prediction step applies to each hypothesis of each track separately
\end{enumerate}

Substituting (\ref{eq:FullDensity}) into (\ref{eq:PGFLPrediction}):
\ifCLASSOPTIONdraftcls
\begin{IEEEeqnarray}{rCl}
G_{t|t-1}[h] &\propto& \exp\{\langle\lambda^\mathrm{b},h\rangle\} G_{t-1|t-1}[1-P^{\mathrm{s}}+P^{\mathrm{s}} p_h] \notag\\
&=& \exp\{\langle\lambda^\mathrm{b},h\rangle\} G^\mathrm{ppp}_{t-1|t-1}[1-P^{\mathrm{s}}+P^{\mathrm{s}} p_h]  
G^\mathrm{mbm}_{t-1|t-1}[1-P^{\mathrm{s}}+P^{\mathrm{s}} p_h] \label{eq:PredictionPGFl}
\end{IEEEeqnarray}
\else
\begin{IEEEeqnarray}{rCl}
G_{t|t-1}[h] &\propto& \exp\{\langle\lambda^\mathrm{b},h\rangle\} G_{t-1|t-1}[1-P^{\mathrm{s}}+P^{\mathrm{s}} p_h] \notag\\
&=& \exp\{\langle\lambda^\mathrm{b},h\rangle\} G^\mathrm{ppp}_{t-1|t-1}[1-P^{\mathrm{s}}+P^{\mathrm{s}} p_h] \notag\\
&& \times G^\mathrm{mbm}_{t-1|t-1}[1-P^{\mathrm{s}}+P^{\mathrm{s}} p_h] \label{eq:PredictionPGFl}
\end{IEEEeqnarray}
\fi
We assign the first two factors in this product to $G^\mathrm{ppp}_{t|t-1}[h]$ and the final to $G^\mathrm{mbm}_{t|t-1}[h]$ to obtain $G_{t|t-1}[h] = G^\mathrm{ppp}_{t|t-1}[h] G^\mathrm{mbm}_{t|t-1}[h]$, where
\begin{align}
G^\mathrm{ppp}_{t|t-1}[h] &\propto \exp\{\langle\lambda^\mathrm{b},h\rangle\} G^\mathrm{ppp}_{t-1|t-1}[1-P^{\mathrm{s}}+P^{\mathrm{s}} p_h] \\
G^\mathrm{mbm}_{t|t-1}[h] &= G^\mathrm{mbm}_{t-1|t-1}[1-P^{\mathrm{s}}+P^{\mathrm{s}} p_h]
\end{align}
We will show that each of these retain their respective forms, \textit{i.e.}\xspace, (\ref{eq:UndetectedTargets}) and (\ref{eq:PreexistingTargets}). $G^\mathrm{ppp}_{t|t-1}[h]$ corresponds directly to the prediction step in the PHD \cite{Mah03}:
\begin{align}
G^\mathrm{ppp}_{t|t-1}[h] 
&\propto \exp\{\langle\lambda^\mathrm{b},h\rangle\} G^\mathrm{ppp}_{t-1|t-1}[1-P^{\mathrm{s}}+P^{\mathrm{s}} p_h] \\
&\propto \exp\{\langle\lambda^\mathrm{b},h\rangle + \langle\lambda^u_{t-1|t-1},1-P^{\mathrm{s}}+P^{\mathrm{s}} p_h\rangle\} \\
&\propto \exp\{\langle\lambda^\mathrm{b},h\rangle + \langle\lambda^u_{t-1|t-1},P^{\mathrm{s}} p_h\rangle\} \\
&= \exp\{\langle\lambda^u_{t|t-1},h\rangle\} 
\end{align}
where $\langle\lambda^u_{t|t-1},h\rangle \triangleq \langle\lambda^\mathrm{b},h\rangle + \langle\lambda^u_{t-1|t-1},P^{\mathrm{s}} p_h\rangle$. Thus $G^\mathrm{ppp}_{t|t-1}[h]$ is indeed a PPP, with intensity given by (\ref{eq:UndetectedPropagation}).

We now consider $G^\mathrm{mbm}_{t|t-1}[h]$; since $G^\mathrm{mbm}_{t-1|t-1}[h]$ is a mixture of MeMBer distributions, this can be calculated via the MeMBer prediction equation\footnote{With target survival/death but without target birth, as this is incorporated in the separate Poisson component $G^\mathrm{ppp}_{t|t-1}[h]$.} \cite[p675]{Mah07}:
\ifCLASSOPTIONdraftcls
\begin{align}
G^\mathrm{mbm}_{t|t-1}[h] &\triangleq G^\mathrm{mbm}_{t-1|t-1}[1-P^{\mathrm{s}}+P^{\mathrm{s}} p_h] \notag\\
&= \sum_a w^a\prod_{i\in{\cal T}_{t-1|t-1}}G_{t-1|t-1}^{i,a^i}[1-P^{\mathrm{s}}+P^{\mathrm{s}} p_h] \\
&= \sum_a w^a\prod_{i\in{\cal T}_{t-1|t-1}}G_{t|t-1}^{i,a^i}[h] \label{eq:PredictionPGFlDecompos}
\end{align}
\else
\begin{align}
\lefteqn{G^\mathrm{mbm}_{t|t-1}[h] \triangleq G^\mathrm{mbm}_{t-1|t-1}[1-P^{\mathrm{s}}+P^{\mathrm{s}} p_h]} \notag\\
&= \sum_a w^a\prod_{i\in{\cal T}_{t-1|t-1}}G_{t-1|t-1}^{i,a^i}[1-P^{\mathrm{s}}+P^{\mathrm{s}} p_h] \\
&= \sum_a w^a\prod_{i\in{\cal T}_{t-1|t-1}}G_{t|t-1}^{i,a^i}[h] \label{eq:PredictionPGFlDecompos}
\end{align}
\fi
where
\ifCLASSOPTIONdraftcls
\begin{IEEEeqnarray}{rCl}
G_{t|t-1}^{i,a^i}[h] &\triangleq& G_{t-1|t-1}^{i,a^i}[1-P^{\mathrm{s}}+P^{\mathrm{s}} p_h] \\
&=& 1-r_{t-1|t-1}^{i,a^i} + r_{t-1|t-1}^{i,a^i}\int{[1-P^{\mathrm{s}}(x')]f_{t-1|t-1}^{i,a^i}(x')\mathrm{d} x'} + \notag\\
&& + r_{t-1|t-1}^{i,a^i}\iint{h(x)f_{t|t-1}(x|x')\mathrm{d} x P^{\mathrm{s}}(x')f_{t-1|t-1}^{i,a^i}(x') \mathrm{d} x'} \\ 
&=& 1-r_{t|t-1}^{i,a^i} + r_{t|t-1}^{i,a^i}\int{h(x)f_{t|t-1}^{i,a^i}(x)\mathrm{d} x} \label{eq:BernoulliPropagation}
\end{IEEEeqnarray}
\else
\begin{IEEEeqnarray}{rCl}
\IEEEeqnarraymulticol{3}{l}{G_{t|t-1}^{i,a^i}[h] \triangleq G_{t-1|t-1}^{i,a^i}[1-P^{\mathrm{s}}+P^{\mathrm{s}} p_h]} \\
&=& 1-r_{t-1|t-1}^{i,a^i} + r_{t-1|t-1}^{i,a^i}\int{[1-P^{\mathrm{s}}(x')]f_{t-1|t-1}^{i,a^i}(x')\mathrm{d} x'} + \notag\\
&& + r_{t-1|t-1}^{i,a^i}\iint{h(x)f_{t|t-1}(x|x')\mathrm{d} x P^{\mathrm{s}}(x')f_{t-1|t-1}^{i,a^i}(x') \mathrm{d} x'} \notag\\ && \\
&=& 1-r_{t|t-1}^{i,a^i} + r_{t|t-1}^{i,a^i}\int{h(x)f_{t|t-1}^{i,a^i}(x)\mathrm{d} x} \label{eq:BernoulliPropagation}
\end{IEEEeqnarray}
\fi
where the final equality results from the definitions of (\ref{eq:ExistProbPropagation}) and (\ref{eq:KinematicDistPropagation}). Recognising (\ref{eq:BernoulliPropagation}) as being Bernoulli, we obtain the result that $G^\mathrm{mbm}_{t|t-1}[h]$ is in the form (\ref{eq:PreexistingTargets}), with $n_{t|t-1}=n_{t-1|t-1}$, $h^i_{t|t-1}=h^i_{t-1|t-1}$, and $w^{i,a^i}_{t|t-1}=w^{i,a^i}_{t-1|t-1}$.
\end{IEEEproof}

\section{Proof of update step}
\label{app:UpdateProof}
The structure of the proof of the Bayes update is as follows:
\begin{enumerate}
\item Substitute the p.g.fl form of the density (\ref{eq:FullDensity}) into the Bayes RFS update formula (\ref{eq:PGFLUpdate}), where the joint p.g.fl of targets and measurements is given by (\ref{eq:JointPGFL})
\item Show that the result is a product of two p.g.fls, where the first is a PPP; this is proven in Lemma~\ref{lem:Structure}
\item Calculate the elements of the update of the second component (which we refer to as the \emph{remaining component})
\begin{enumerate}
\item Updating the detected portion of a PPP with PPP-distributed clutter (Lemma~\ref{lem:PoisUpdate})
\item Updating a Bernoulli component (Lemma~\ref{lem:BernoulliUpdate})
\end{enumerate}
Both of these elements use the simple result in Lemma~\ref{lem:ScaledBernoulli}. The complete update of the remaining component can then be calculated from these elements using the product rule (\ref{eq:ProductRule}).
\item By linearity of the derivative operator, the update in the remaining component can be broken into a sum with a term for each global hypothesis $a$ in the prior MBM $G^\mathrm{mbm}_{t|t-1}[h]$ (the first step in Theorem~\ref{th:Update}), effectively computing the update separately for each prior association hypothesis
\item Given a prior global hypothesis (\textit{i.e.}\xspace, a global association history hypothesis involving measurements from previous time steps), the update can be calculated using the product rule (\ref{eq:ProductRule}); the resulting structure is shown in Lemma~\ref{lem:ConvolutionEquiv} to be in a form similar to (\ref{eq:PreexistingTargets}) with weights decomposing according to (\ref{eq:GlobalHypWeight})
\item Subsequently (in Theorem~\ref{th:Update}) the sum over prior association hypotheses and hypotheses in the new scan are combined into a single sum
\end{enumerate}

\begin{lemma}\label{lem:Structure}
The updated distribution $G_{t|t}[h]$ has the form
\begin{equation}\label{eq:UpdateRetainsStructure}
G_{t|t}[h] = G^\mathrm{ppp}_{t|t}[h] G^\mathrm{mbm}_{t|t}[h]
\end{equation}
where
\ifCLASSOPTIONdraftcls
\begin{IEEEeqnarray}{rCl}
G^\mathrm{ppp}_{t|t}[h] &\propto& \exp\{\langle\lambda_{t|t}^u,h\rangle\} \label{eq:UpdateUndetectedPGFl} \\
G^\mathrm{mbm}_{t|t}[h] &\propto& \frac{\delta}{\delta Z_t}\big\{ \exp\{\langle\lambda^\mathrm{fa},g\rangle + \langle\lambda^u_{t|t-1},h P^{\mathrm{d}} p_g\rangle\}
\cdot G^\mathrm{mbm}_{t|t-1}[h(1-P^{\mathrm{d}} + P^{\mathrm{d}} p_g)] \big\}\bigg|_{g=0} \label{eq:UpdateMBMPGFl}
\end{IEEEeqnarray}
\else
\begin{IEEEeqnarray}{rCl}
G^\mathrm{ppp}_{t|t}[h] &\propto& \exp\{\langle\lambda_{t|t}^u,h\rangle\} \label{eq:UpdateUndetectedPGFl} \\
G^\mathrm{mbm}_{t|t}[h] &\propto& \frac{\delta}{\delta Z_t}\big\{ \exp\{\langle\lambda^\mathrm{fa},g\rangle + \langle\lambda^u_{t|t-1},h P^{\mathrm{d}} p_g\rangle\} \notag\\
&& \times G^\mathrm{mbm}_{t|t-1}[h(1-P^{\mathrm{d}} + P^{\mathrm{d}} p_g)] \big\}\bigg|_{g=0} \label{eq:UpdateMBMPGFl}
\end{IEEEeqnarray}
\fi
and $\lambda_{t|t}^u(x)$ is given in (\ref{eq:UndetectedUpdate}).
\end{lemma}

\begin{IEEEproof} %[Proof of \lemref{lem:Structure}]
Substituting (\ref{eq:JointPGFL}) and (\ref{eq:FullDensity}) into (\ref{eq:PGFLUpdate}),
\ifCLASSOPTIONdraftcls
\begin{IEEEeqnarray}{rCl}
G_{t|t}[h] &\propto& \frac{\delta}{\delta Z_t}\big\{ \exp\{\langle\lambda^\mathrm{fa},g\rangle\} \cdot G_{t|t-1}[h(1-P^{\mathrm{d}} + P^{\mathrm{d}} p_g)]\big\}\bigg|_{g=0} \label{eq:UpdateStructure1}\\
&=& \frac{\delta}{\delta Z_t}\big\{ \exp\{\langle\lambda^\mathrm{fa},g\rangle\} \cdot G^\mathrm{ppp}_{t|t-1}[h(1-P^{\mathrm{d}} + P^{\mathrm{d}} p_g)] % \cdot\notag\\* &&
 \cdot G^\mathrm{mbm}_{t|t-1}[h(1-P^{\mathrm{d}} + P^{\mathrm{d}} p_g)] \big\}\bigg|_{g=0} \label{eq:UpdateStructure2}\\
&\propto& \frac{\delta}{\delta Z_t}\big\{ \exp\{\langle\lambda^\mathrm{fa},g\rangle + \langle\lambda^u_{t|t-1},h(1-P^{\mathrm{d}} + P^{\mathrm{d}} p_g)\rangle\} %\cdot\notag\\* &&
 \cdot G^\mathrm{mbm}_{t|t-1}[h(1-P^{\mathrm{d}} + P^{\mathrm{d}} p_g)] \big\}\bigg|_{g=0} \label{eq:UpdateStructure3}\\
&=& \exp\{\langle\lambda^u_{t|t-1},h(1-P^{\mathrm{d}})\rangle\} %\cdot\notag\\* &&
 \cdot \frac{\delta}{\delta Z_t}\big\{ \exp\{\langle\lambda^\mathrm{fa},g\rangle + \langle\lambda^u_{t|t-1},h P^{\mathrm{d}} p_g\rangle\} %\cdot\notag\\* &&
 \cdot G^\mathrm{mbm}_{t|t-1}[h(1-P^{\mathrm{d}} + P^{\mathrm{d}} p_g)] \big\}\bigg|_{g=0} \label{eq:UpdateStructure4}
\label{eq:UpdateStructure}
\end{IEEEeqnarray}
\else
\begin{IEEEeqnarray}{rCl}
\IEEEeqnarraymulticol{3}{l}{G_{t|t}[h]} \notag \\
&\propto& \frac{\delta}{\delta Z_t}\big\{ \exp\{\langle\lambda^\mathrm{fa},g\rangle\} G_{t|t-1}[h(1-P^{\mathrm{d}} + P^{\mathrm{d}} p_g)]\big\}\bigg|_{g=0} \label{eq:UpdateStructure1}\\
&=& \frac{\delta}{\delta Z_t}\big\{ \exp\{\langle\lambda^\mathrm{fa},g\rangle\} G^\mathrm{ppp}_{t|t-1}[h(1-P^{\mathrm{d}} + P^{\mathrm{d}} p_g)] \notag\\* 
&& \times G^\mathrm{mbm}_{t|t-1}[h(1-P^{\mathrm{d}} + P^{\mathrm{d}} p_g)] \big\}\bigg|_{g=0} \label{eq:UpdateStructure2}\\
&\propto& \frac{\delta}{\delta Z_t}\big\{ \exp\{\langle\lambda^\mathrm{fa},g\rangle + \langle\lambda^u_{t|t-1},h(1-P^{\mathrm{d}} + P^{\mathrm{d}} p_g)\rangle\}\notag\\*
&& \times G^\mathrm{mbm}_{t|t-1}[h(1-P^{\mathrm{d}} + P^{\mathrm{d}} p_g)] \big\}\bigg|_{g=0} \label{eq:UpdateStructure3}\\
&=& \exp\{\langle\lambda^u_{t|t-1},h(1-P^{\mathrm{d}})\rangle\}\notag\\*
&& \times \frac{\delta}{\delta Z_t}\big\{ \exp\{\langle\lambda^\mathrm{fa},g\rangle + \langle\lambda^u_{t|t-1},h P^{\mathrm{d}} p_g\rangle\}\notag\\*
&& \times G^\mathrm{mbm}_{t|t-1}[h(1-P^{\mathrm{d}} + P^{\mathrm{d}} p_g)] \big\}\bigg|_{g=0} \label{eq:UpdateStructure4}
\label{eq:UpdateStructure}
\end{IEEEeqnarray}
\fi
where (\ref{eq:UpdateStructure1}) substitutes (\ref{eq:JointPGFL}) into (\ref{eq:PGFLUpdate}), (\ref{eq:UpdateStructure2}) substitutes (\ref{eq:FullDensity}) for $G_{t|t-1}[\cdot]$, (\ref{eq:UpdateStructure3}) substitutes (\ref{eq:UndetectedTargets}) for $G^\mathrm{ppp}_{t|t-1}[h]$, and (\ref{eq:UpdateStructure4}) exploits the fact that 
\ifCLASSOPTIONdraftcls
\begin{IEEEeqnarray}{rCl}
\langle\lambda^u_{t|t-1},h(1-P^{\mathrm{d}} + P^{\mathrm{d}} p_g)\rangle &=& \int{ \lambda^u_{t|t-1}(x) h(x) \{1-P^{\mathrm{d}}(x) + P^{\mathrm{d}}(x) p_g(x)\}\mathrm{d} x} \\
&=& \int{ \lambda^u_{t|t-1}(x) h(x) \{1-P^{\mathrm{d}}(x)\} \mathrm{d} x} %\notag\\ &&
 + \int{ \lambda^u_{t|t-1}(x) h(x) P^{\mathrm{d}}(x) p_g(x)\mathrm{d} x} \\
&=& \langle\lambda^u_{t|t-1},h(1-P^{\mathrm{d}})\rangle + \langle\lambda^u_{t|t-1},h(P^{\mathrm{d}} p_g)\rangle
\end{IEEEeqnarray}
\else
\begin{IEEEeqnarray}{rCl}
\IEEEeqnarraymulticol{3}{l}{\langle\lambda^u_{t|t-1},h(1-P^{\mathrm{d}} + P^{\mathrm{d}} p_g)\rangle} \notag\\
&=& \int{ \lambda^u_{t|t-1}(x) h(x) \{1-P^{\mathrm{d}}(x) + P^{\mathrm{d}}(x) p_g(x)\}\mathrm{d} x} \\
&=& \int{ \lambda^u_{t|t-1}(x) h(x) \{1-P^{\mathrm{d}}(x)\} \mathrm{d} x} \notag\\
&& + \int{ \lambda^u_{t|t-1}(x) h(x) P^{\mathrm{d}}(x) p_g(x)\mathrm{d} x} \\
&=& \langle\lambda^u_{t|t-1},h(1-P^{\mathrm{d}})\rangle + \langle\lambda^u_{t|t-1},h(P^{\mathrm{d}} p_g)\rangle
\end{IEEEeqnarray}
\fi
and that the factor $\exp\{\langle\lambda^u_{t|t-1},h(1-P^{\mathrm{d}})\rangle\}$ is independent of the functional of differentiation $g$, and thus can be taken outside the derivative. Thus, by setting $G^\mathrm{ppp}_{t|t}[h]$ to be the first line in (\ref{eq:UpdateStructure4}), and $G^\mathrm{mbm}_{t|t}[h]$ to be equal to the second and third lines, we arrive at the results in (\ref{eq:UndetectedUpdate}), (\ref{eq:UpdateRetainsStructure}), (\ref{eq:UpdateUndetectedPGFl}) and (\ref{eq:UpdateMBMPGFl}).
\end{IEEEproof}

\begin{lemma}\label{lem:ScaledBernoulli}
Any p.g.fl of the following form\footnote{with $a\geq 0$, $b(x)\geq 0$, $0\leq \int{b(x)\mathrm{d} x} < \infty$. If $\int{b(x)\mathrm{d} x}=0$ then any choice of $f(x)$ will satisfy (\ref{eq:ScaledBernoulli}). If $a=0$ and $\int{b(x)\mathrm{d} x}=0$ then any choice of $r$ and $f(x)$ will suffice.} is an unnormalised Bernoulli distribution:
\begin{equation}\label{eq:ScaledBernoulli}
a + \langle b,h\rangle = w (1-r + r \cdot \langle f,h\rangle)
\end{equation}
where $w = a + \int{b(x)\mathrm{d} x} = a + \langle b,1\rangle$, $r = \frac{\langle b,1\rangle}{a+\langle b,1\rangle}$ and $f(x) = \frac{b(x)}{\langle b,1\rangle}$.
\end{lemma}
The proof of Lemma~\ref{lem:ScaledBernoulli} follows immediately from substituting $w$, $r$ and $f(x)$ into the RHS of (\ref{eq:ScaledBernoulli}). Lemma~\ref{lem:PoisUpdate} calculates the unnormalised Bayes update for an unnormalised Poisson distribution in Poisson clutter (note that the undetected portion of the Poisson distribution has been removed as it is represented separately, in $G^\mathrm{ppp}_{t|t}[h]$).
\begin{lemma}\label{lem:PoisUpdate}
\ifCLASSOPTIONdraftcls
\begin{equation}
\frac{\delta}{\delta Z} \exp\{\langle\lambda^\mathrm{fa},g\rangle + \langle\lambda^u_{t|t-1},h P^{\mathrm{d}} p_g\rangle\}\bigg|_{g=0} 
= \prod_{z\in Z} w_{t|t}^z (1-r_{t|t}^z + r_{t|t}^z \langle f_{t|t}^z,h\rangle) 
\end{equation}
\else
\begin{multline}
\frac{\delta}{\delta Z} \exp\{\langle\lambda^\mathrm{fa},g\rangle + \langle\lambda^u_{t|t-1},h P^{\mathrm{d}} p_g\rangle\}\bigg|_{g=0} \\
= \prod_{z\in Z} w_{t|t}^z (1-r_{t|t}^z + r_{t|t}^z \langle f_{t|t}^z,h\rangle) 
\end{multline}
\fi
where $w_{t|t}^z$, $r_{t|t}^z$ and $f_{t|t}^z(x)$ are given in (\ref{eq:PoisUpdateW}), (\ref{eq:PoisUpdatePex}) and (\ref{eq:PoisUpdateKin}) respectively (replacing $z$ with $z_t^j$).
\end{lemma}
\begin{IEEEproof}
We use the derivative of a linear functional, and the chain rule: \cite[p395]{Mah07}
\begin{align}
\frac{\delta}{\delta z}\langle f,g\rangle =& f(z) \label{eq:LinearFunctional} \\
\frac{\delta}{\delta z} f(F[g]) &= 
\frac{\delta}{\delta z}F[g] \cdot \left.\frac{\mathrm{d}}{\mathrm{d} y} f(y)\right|_{y=F[g]} \label{eq:ChainRule}
\end{align}
Using the result for linear functionals (\ref{eq:LinearFunctional}):
\ifCLASSOPTIONdraftcls
\begin{equation}
\frac{\delta}{\delta z} \left( \langle\lambda^\mathrm{fa},g\rangle + \langle\lambda^u_{t|t-1},h P^{\mathrm{d}} p_g\rangle \right) 
= \lambda^\mathrm{fa}(z) + \langle\lambda^u_{t|t-1},h P^{\mathrm{d}} f(z|\cdot)\rangle
\end{equation}
\else
\begin{multline}
\frac{\delta}{\delta z} \left( \langle\lambda^\mathrm{fa},g\rangle + \langle\lambda^u_{t|t-1},h P^{\mathrm{d}} p_g\rangle \right) \\
= \lambda^\mathrm{fa}(z) + \langle\lambda^u_{t|t-1},h P^{\mathrm{d}} f(z|\cdot)\rangle
\end{multline}
\fi
Subsequently applying the chain rule (\ref{eq:ChainRule}):
\ifCLASSOPTIONdraftcls
\begin{equation}
\frac{\delta}{\delta z} \exp\{ \langle\lambda^\mathrm{fa},g\rangle + \langle\lambda^u_{t|t-1},h P^{\mathrm{d}} p_g\rangle \} 
= \exp\{ \langle\lambda^\mathrm{fa},g\rangle + \langle\lambda^u_{t|t-1},h P^{\mathrm{d}} p_g\rangle \} 
\cdot ( \lambda^\mathrm{fa}(z) + \langle\lambda^u_{t|t-1},h P^{\mathrm{d}} f(z|\cdot)\rangle )
\end{equation}
\else
\begin{multline}
\frac{\delta}{\delta z} \exp\{ \langle\lambda^\mathrm{fa},g\rangle + \langle\lambda^u_{t|t-1},h P^{\mathrm{d}} p_g\rangle \} \\
= \exp\{ \langle\lambda^\mathrm{fa},g\rangle + \langle\lambda^u_{t|t-1},h P^{\mathrm{d}} p_g\rangle \}  \\
\times ( \lambda^\mathrm{fa}(z) + \langle\lambda^u_{t|t-1},h P^{\mathrm{d}} f(z|\cdot)\rangle )
\end{multline}
\fi
Iterating the derivative, we find
\ifCLASSOPTIONdraftcls
\begin{equation}
\frac{\delta}{\delta Z} \exp\{ \langle\lambda^\mathrm{fa},g\rangle + \langle\lambda^u_{t|t-1},h P^{\mathrm{d}} p_g\rangle \} 
= \exp\{ \langle\lambda^\mathrm{fa},g\rangle + \langle\lambda^u_{t|t-1},h P^{\mathrm{d}} p_g\rangle \} 
\cdot \prod_{z\in Z}\left( \lambda^\mathrm{fa}(z) + \langle\lambda^u_{t|t-1},h P^{\mathrm{d}} f(z|\cdot)\rangle \right)
\end{equation}
\else
\begin{multline}
\frac{\delta}{\delta Z} \exp\{ \langle\lambda^\mathrm{fa},g\rangle + \langle\lambda^u_{t|t-1},h P^{\mathrm{d}} p_g\rangle \} \\
= \exp\{ \langle\lambda^\mathrm{fa},g\rangle + \langle\lambda^u_{t|t-1},h P^{\mathrm{d}} p_g\rangle \}  \\
\times \prod_{z\in Z}\left( \lambda^\mathrm{fa}(z) + \langle\lambda^u_{t|t-1},h P^{\mathrm{d}} f(z|\cdot)\rangle \right)
\end{multline}
\fi
Setting $g=0$, and applying Lemma~\ref{lem:ScaledBernoulli}, we obtain the desired result.
\end{IEEEproof}
Lemma~\ref{lem:BernoulliUpdate} calculates the p.g.fl form of the Bayes update for an unnormalised Bernoulli distribution.
\begin{lemma}\label{lem:BernoulliUpdate}
\ifCLASSOPTIONdraftcls
\begin{align}
\frac{\delta}{\delta Z}& w_{t|t-1}^{i,\tilde{a}^i} \Big( 1-r_{t|t-1}^{i,\tilde{a}^i} 
 + r_{t|t-1}^{i,\tilde{a}^i}\langle f_{t|t-1}^{i,\tilde{a}^i},h(1-P^{\mathrm{d}} + P^{\mathrm{d}} p_g)\rangle\Big)\bigg|_{g=0}\notag \\
=& \begin{cases}
w^{i,\tilde{a}^i}_{t|t}(1-r^{i,\tilde{a}^i}_{t|t}+r^{i,\tilde{a}^i}_{t|t}\langle f^{i,\tilde{a}^i}_{t|t},h\rangle), & Z=\emptyset \\
w^{i,a^i}_{t|t}(1-r^{i,a^i}_{t|t}+r^{i,a^i}_{t|t}\langle f^{i,a^i}_{t|t},h\rangle), & Z=\{z_t^j\} \\
0, & \mbox{otherwise}
\end{cases}
\end{align}
\else
\begin{align}
\frac{\delta}{\delta Z}& w_{t|t-1}^{i,\tilde{a}^i} \Big( 1-r_{t|t-1}^{i,\tilde{a}^i} + \notag\\
& + r_{t|t-1}^{i,\tilde{a}^i}\langle f_{t|t-1}^{i,\tilde{a}^i},h(1-P^{\mathrm{d}} + P^{\mathrm{d}} p_g)\rangle\Big)\bigg|_{g=0}\notag \\
=& \begin{cases}
w^{i,\tilde{a}^i}_{t|t}(1-r^{i,\tilde{a}^i}_{t|t}+r^{i,\tilde{a}^i}_{t|t}\langle f^{i,\tilde{a}^i}_{t|t},h\rangle), & Z=\emptyset \\
w^{i,a^i}_{t|t}(1-r^{i,a^i}_{t|t}+r^{i,a^i}_{t|t}\langle f^{i,a^i}_{t|t},h\rangle), & Z=\{z_t^j\} \\
0, & \mbox{otherwise}
\end{cases}
\end{align}
\fi
where $w^{i,\tilde{a}^i}_{t|t}$, $r^{i,\tilde{a}^i}_{t|t}$ and $f^{i,\tilde{a}^i}_{t|t}(x)$ are given in (\ref{eq:MissUpdateW}), (\ref{eq:MissUpdatePex}) and (\ref{eq:MissUpdateKin}) respectively, and $w^{i,a^i}_{t|t}$, $r^{i,a^i}_{t|t}$ and $f^{i,a^i}_{t|t}(x)$ are given in (\ref{eq:DetUpdateW}), (\ref{eq:DetUpdatePex}) and (\ref{eq:DetUpdateKin}) respectively.
\end{lemma}
\begin{IEEEproof}
The case with $Z=\emptyset$ results simply from Lemma~\ref{lem:ScaledBernoulli}. When $Z=\{z_t^j\}$, we observe that:
\ifCLASSOPTIONdraftcls
\begin{equation}
\langle  f_{t|t-1}^{i,\tilde{a}^i},h(1-P^{\mathrm{d}} + P^{\mathrm{d}} p_g)\rangle 
= \langle f_{t|t-1}^{i,\tilde{a}^i},h(1-P^{\mathrm{d}})\rangle + 
\langle f_{t|t-1}^{i,\tilde{a}^i},hP^{\mathrm{d}} p_g\rangle
\end{equation}
\else
\begin{multline}
\langle  f_{t|t-1}^{i,\tilde{a}^i},h(1-P^{\mathrm{d}} + P^{\mathrm{d}} p_g)\rangle \\
= \langle f_{t|t-1}^{i,\tilde{a}^i},h(1-P^{\mathrm{d}})\rangle + 
\langle f_{t|t-1}^{i,\tilde{a}^i},hP^{\mathrm{d}} p_g\rangle
\end{multline}
\fi
where
\ifCLASSOPTIONdraftcls
\begin{align}
\langle f_{t|t-1}^{i,\tilde{a}^i},hP^{\mathrm{d}} p_g\rangle 
&= \int f_{t|t-1}^{i,\tilde{a}^i}(x)h(x)P^{\mathrm{d}}(x)\int f(z|x)g(z)\mathrm{d} z\mathrm{d} x \\
&= \int g(z)\int h(x)f_{t|t-1}^{i,\tilde{a}^i}(x)P^{\mathrm{d}}(x)f(z|x)\mathrm{d} x\mathrm{d} z
\end{align}
\else
\begin{align}
\langle & f_{t|t-1}^{i,\tilde{a}^i},hP^{\mathrm{d}} p_g\rangle \notag\\
&= \int f_{t|t-1}^{i,\tilde{a}^i}(x)h(x)P^{\mathrm{d}}(x)\int f(z|x)g(z)\mathrm{d} z\mathrm{d} x \\
&= \int g(z)\int h(x)f_{t|t-1}^{i,\tilde{a}^i}(x)P^{\mathrm{d}}(x)f(z|x)\mathrm{d} x\mathrm{d} z
\end{align}
\fi
Subsequently, we apply (\ref{eq:LinearFunctional}) and Lemma~\ref{lem:ScaledBernoulli}.
\end{IEEEproof}

Lemma~\ref{lem:ConvolutionEquiv} shows that the product rule (\ref{eq:ProductRule}) can be expressed equivalently in a form similar to (\ref{eq:PreexistingTargets}). This equivalence is the point at which data association arises in the derivation; the terms of this sum can be interpreted as global association hypotheses. When we employ Lemma~\ref{lem:ConvolutionEquiv}, $F_0^Z[g]$ represents measurements arising from false alarms or previously unknown targets, while $F_i^Z[g]$, $i>0$ represents measurements from pre-existing tracks.

\begin{lemma}\label{lem:ConvolutionEquiv}
Suppose we are given set-parameterised functionals $F_i^Z[g]=\frac{\delta F_i}{\delta Z}[g]$, $i\in\{0,...,n\}$, such that $F_0^Z[g]=\prod_{z\in Z}f_0^z[g]$ and if $i\geq 1$ and $|Z|\geq 2$ then $F_i^Z[g]=0$. If $Z = \{z^1,\dots,z^{m}\}$, then
\begin{equation}
\sum_{W_0 \uplus \cdots \uplus W_n=Z}\prod_{i=0}^n F_i^{W_i}[g]
= \sum_{\bar{a}\in{\cal A}_t}\prod_{i=1}^{n+m} F_i^{Z^{\bar{a}^i}}[g]
\end{equation}
where
\ifCLASSOPTIONdraftcls
\begin{multline}
{\cal A}_t = \bigg\{
(\bar{a}^1,\dots,\bar{a}^{n+m})
\bigg| 
\bar{a}^i\in\{0,\dots,m\},\mbox{ for } i\in\{1,\dots,n\}, 
\bar{a}^{n+j}\in\{0,j\},\mbox{ for } j\in\{1,\dots,m\}, \\
{\textstyle\bigcup_{i=1}^{n+m}}{\cal M}_t(i,\bar{a}^i)=\{(t,1),\dots,(t,m)\},
{\cal M}^t(i,\bar{a}^i)\cap{\cal M}^t(i',\bar{a}^{i'})=\emptyset \;\forall\; i\neq i' 
\bigg\}
\end{multline}
\else
\begin{multline}
{\cal A}_t = \bigg\{
(\bar{a}^1,\dots,\bar{a}^{n+m})
\bigg| 
\bar{a}^i\in\{0,\dots,m\},\mbox{ for } i\in\{1,\dots,n\}, \\
\bar{a}^{n+j}\in\{0,j\},\mbox{ for } j\in\{1,\dots,m\}, \\
{\textstyle\bigcup_{i=1}^{n+m}}{\cal M}_t(i,\bar{a}^i)=\{(t,1),\dots,(t,m)\},
\\
{\cal M}^t(i,\bar{a}^i)\cap{\cal M}^t(i',\bar{a}^{i'})=\emptyset \;\forall\; i\neq i' 
\bigg\}
\end{multline}
\fi
with ${\cal M}_t(i,0)=\emptyset$ and ${\cal M}_t(i,j)=\{(t,j)\}$. Finally,
\begin{equation}
Z^{\bar{a}^i} \triangleq \begin{cases}
\emptyset, & \bar{a}^i = 0 \\
z^{\bar{a}^i}, & \bar{a}^i \in\{1,\dots,m\}
\end{cases}
\end{equation}
and for $j\in\{1,\dots,m\}$,
\begin{equation}
F_{n+j}^Z[g] \triangleq \begin{cases}
1, & Z=\emptyset \\
f_0^{z^j}[g], & Z=\{z^j\} \\
0, & \mbox{otherwise}
\end{cases} 
\end{equation}
\end{lemma}
\begin{IEEEproof}
The result follows from the construction of ${\cal A}_t$ and the other components. To observe the correspondence of LHS and RHS terms, take $\bar{a}=(\bar{a}^1,\dots,\bar{a}^{n+m})\in{\cal A}_t$. Set $W_0=\bigcup_{i=n+1}^{n+m}{\cal M}_t(i,\bar{a}^i)$, and for $i\in\{1,\dots,n\}$ set $W_i={\cal M}_t(i,a^i)$. By inspection, the corresponding LHS and RHS terms are equivalent. Similarly, take any non-zero LHS term, and for $i\in\{1,\dots,n\}$ set
\begin{equation}
\bar{a}^i = \begin{cases}
0, & W^i = \emptyset \\
j, & W^i = \{z^j\} 
\end{cases}
\end{equation}
Similarly, for $j\in\{1,\dots,m\}$, set
\begin{equation}
\bar{a}^{n+j} = \begin{cases}
0, & z_j\notin W_0 \\
j, & z_j\in W_0
\end{cases}
\end{equation}
Again, by inspection, these terms will be in correspondence.
\end{IEEEproof}

With these preliminary results, we are now ready to prove the main theorem.

\begin{IEEEproof}[Proof of Theorem~\ref{th:Update}]
Following Lemma~\ref{lem:Structure}, the necessary result is that  $G^\mathrm{mbm}_{t|t}[h]$ in (\ref{eq:UpdateMBMPGFl}) is made equal (or proportional) to (\ref{eq:PreexistingTargets}) through the equalities made in the statement of the theorem (\textit{i.e.}\xspace, that $G^\mathrm{mbm}_{t|t}[h]$ is a mixture of multi-Bernoulli distributions). Expanding the form of $G^\mathrm{mbm}_{t|t-1}[h]$ using (\ref{eq:PreexistingTargets}) and (\ref{eq:GlobalHypWeight}) and applying the linearity property of differentials, we can write:
\ifCLASSOPTIONdraftcls
\begin{equation}\label{eq:MBMUpdate1}
G^\mathrm{mbm}_{t|t}[h] \propto
\sum_{\tilde{a}\in{\cal A}^{t|t-1}} \frac{\delta}{\delta Z_t}\bigg( 
\exp\{\langle\lambda^\mathrm{fa},g\rangle + \langle\lambda^u_{t|t-1},h P^{\mathrm{d}} p_g\rangle\} 
\cdot \prod_{i\in{\cal T}_{t|t-1}}w^{i,\tilde{a}^i}_{t|t-1}G^{i,\tilde{a}^i}_{t|t-1}[h(1-P^{\mathrm{d}} + P^{\mathrm{d}} p_g)]\bigg)\Bigg|_{g=0}
\end{equation}
\else
\begin{multline}\label{eq:MBMUpdate1}
G^\mathrm{mbm}_{t|t}[h] \propto
\sum_{\tilde{a}\in{\cal A}^{t|t-1}} \frac{\delta}{\delta Z_t}\bigg( 
\exp\{\langle\lambda^\mathrm{fa},g\rangle + \langle\lambda^u_{t|t-1},h P^{\mathrm{d}} p_g\rangle\} \\
\times \prod_{i\in{\cal T}_{t|t-1}}w^{i,\tilde{a}^i}_{t|t-1}G^{i,\tilde{a}^i}_{t|t-1}[h(1-P^{\mathrm{d}} + P^{\mathrm{d}} p_g)]\bigg)\Bigg|_{g=0}
\end{multline}
\fi
Considering a single term in this sum (\textit{i.e.}\xspace, a particular choice of $\tilde{a}$), we apply the product rule (\ref{eq:ProductRule}), taking derivatives with respect to the functional $g$, setting $n=n_{t|t-1}$, and
\begin{align}
F_{\tilde{a},0}[g,h] &= \exp\{\langle\lambda^\mathrm{fa},g\rangle + \langle\lambda^u_{t|t-1},h P^{\mathrm{d}} p_g\rangle\} \\
F_{\tilde{a},i}[g,h] &= w^{i,\tilde{a}^i}_{t|t-1}G^{i,\tilde{a}^i}_{t|t-1}[h(1-P^{\mathrm{d}} + P^{\mathrm{d}} p_g)], \; i>0
\end{align}
to obtain
\ifCLASSOPTIONdraftcls
\begin{multline}
\frac{\delta}{\delta Z_t}\bigg(\exp\{\langle\lambda^\mathrm{fa},g\rangle + \langle\lambda^u_{t|t-1},h P^{\mathrm{d}} p_g\rangle\} 
\cdot \prod_{i\in{\cal T}_{t|t-1}}w^{i,\tilde{a}^i}_{t|t-1}G^{i,\tilde{a}^i}_{t|t-1}[h(1-P^{\mathrm{d}} + P^{\mathrm{d}} p_g)]\bigg) \\
=
\sum_{W_0 \uplus \cdots \uplus W_n = Z_t}
\frac{\delta F_{\tilde{a},0}}{\delta W_0}[g,h] \cdots
\frac{\delta F_{\tilde{a},n}}{\delta W_n}[g,h]
\end{multline}
\else
\begin{multline}
\frac{\delta}{\delta Z_t}\bigg(\exp\{\langle\lambda^\mathrm{fa},g\rangle + \langle\lambda^u_{t|t-1},h P^{\mathrm{d}} p_g\rangle\}  \\
\times \prod_{i\in{\cal T}_{t|t-1}}w^{i,\tilde{a}^i}_{t|t-1}G^{i,\tilde{a}^i}_{t|t-1}[h(1-P^{\mathrm{d}} + P^{\mathrm{d}} p_g)]\bigg) \\
=
\sum_{W_0 \uplus \cdots \uplus W_n = Z_t}
\frac{\delta F_{\tilde{a},0}}{\delta W_0}[g,h] \cdots
\frac{\delta F_{\tilde{a},n}}{\delta W_n}[g,h]
\end{multline}
\fi
Subsequently applying Lemma~\ref{lem:ConvolutionEquiv}, we obtain:
\ifCLASSOPTIONdraftcls
\begin{multline}
\frac{\delta}{\delta Z_t}\bigg(\exp\{\langle\lambda^\mathrm{fa},g\rangle + \langle\lambda^u_{t|t-1},h P^{\mathrm{d}} p_g\rangle\} %\cdot \\
\cdot \prod_{i\in{\cal T}_{t|t-1}}w^{i,\tilde{a}^i}_{t|t-1}G^{i,\tilde{a}^i}_{t|t-1}[h(1-P^{\mathrm{d}} + P^{\mathrm{d}} p_g)]\bigg) \\
= \sum_{\bar{a}\in{\cal A}_t}\prod_{i=1}^{n_{t|t-1}+m_t} F_{\tilde{a},i}^{Z^{\bar{a}^i}}[g,h]
\end{multline}
\else
\begin{multline}
\frac{\delta}{\delta Z_t}\bigg(\exp\{\langle\lambda^\mathrm{fa},g\rangle + \langle\lambda^u_{t|t-1},h P^{\mathrm{d}} p_g\rangle\}  \\
\times \prod_{i\in{\cal T}_{t|t-1}}w^{i,\tilde{a}^i}_{t|t-1}G^{i,\tilde{a}^i}_{t|t-1}[h(1-P^{\mathrm{d}} + P^{\mathrm{d}} p_g)]\bigg) \\
= \sum_{\bar{a}\in{\cal A}_t}\prod_{i=1}^{n_{t|t-1}+m_t} F_{\tilde{a},i}^{Z^{\bar{a}^i}}[g,h]
\end{multline}
\fi
Substituting this into (\ref{eq:MBMUpdate1}), we obtain
\begin{equation}
G^\mathrm{mbm}_{t|t}[h] \propto \sum_{\tilde{a}\in{\cal A}^{t|t-1},\bar{a}\in{\cal A}_t}\prod_{i=1}^{n_{t|t-1}+m_t} F_{\tilde{a},i}^{Z^{\bar{a}^i}}[g,h]\bigg|_{g=0}
\end{equation}
Observing that the set of events $a\in{\cal A}^{t|t}$ covers the same events as $(\tilde{a}\in{\cal A}^{t|t-1},\bar{a}\in{\cal A}_t)$, we arrive at our final result
\begin{equation}
G^\mathrm{mbm}_{t|t}[h] \propto \sum_{a\in{\cal A}^{t|t}} \prod_{i\in{\cal T}_{t|t}}w^{i,a^i}_{t|t}G_{t|t}^{i,a^i}[h]
\end{equation}
where, for the term $a\in{\cal A}^{t|t}$ corresponding to $\tilde{a}\in{\cal A}^{t|t-1}$ and $\bar{a}\in{\cal A}_t$, (matching terms are identified by ${\cal M}^{t}(i,a^i)={\cal M}^{t-1}(i,\tilde{a}^i)\cup{\cal M}_t(i,\bar{a}^i)$)
\ifCLASSOPTIONdraftcls
\begin{equation}
w^{i,a^i}_{t|t}G_{t|t}^{i,a^i}[h] = F_{\tilde{a},i}^{Z^{\bar{a}^i}}[g,h]\bigg|_{g=0} 
= w_{t|t}^{i,a^i}(1-r^{i,a^i}_{t|t} + r^{i,a^i}_{t|t}\langle f^{i,a^i}_{t|t},h\rangle)
\end{equation}
\else
\begin{align}
w^{i,a^i}_{t|t}G_{t|t}^{i,a^i}[h] &= F_{\tilde{a},i}^{Z^{\bar{a}^i}}[g,h]\bigg|_{g=0} \\
&= w_{t|t}^{i,a^i}(1-r^{i,a^i}_{t|t} + r^{i,a^i}_{t|t}\langle f^{i,a^i}_{t|t},h\rangle)
\end{align}
\fi
Lemmas \ref{lem:PoisUpdate} and \ref{lem:BernoulliUpdate} confirm that the definitions in the statement of Theorem~\ref{th:Update} achieve this equality with the respective parameters given in (\ref{eq:MissUpdateW})-(\ref{eq:MissUpdateKin}) (missed detections on existing tracks, \textit{i.e.}\xspace, $i\in\{1,\dots,n_{t|t-1}\}$, $a^i\in\{1,\dots,h_{t|t-1}^i\}$), (\ref{eq:DetUpdateW})-(\ref{eq:DetUpdateKin}) (detections updating existing tracks, \textit{i.e.}\xspace, $i\in\{1,\dots,n_{t|t-1}\}$, $a^i\in\{h_{t|t-1}^i+1,\dots,h_{t|t}^i\}$), (\ref{eq:NewTargetNonExistWQ}) (new tracks without measurements, \textit{i.e.}\xspace, $i\in\{n_{t|t-1}+1,\dots,n_{t|t-1}+m_t\}$, $a^i=1$), and (\ref{eq:PoisUpdateW})-(\ref{eq:PoisUpdateKin}) (updates of new tracks, \textit{i.e.}\xspace, $i\in\{n_{t|t-1}+1,\dots,n_{t|t-1}+m_t\}$, $a^i=2$).
\end{IEEEproof}

\section{Implementation pseudo-code}
\label{app:Pseudocode}
{\noindent}The simplest implementation of the algorithm uses a Gaussian approximation of each track, and a Gaussian mixture approximation of the PPP intensity of unknown targets. Pseudo-code for this is provided in this section for completeness; an earlier variant appeared in \cite{Wil11b} along with experimental results. Note that the experiments in Section~\ref{sec:Results} utilised a Gaussian mixture implementation, which is complicated by hypothesis management, \textit{etc}\xspace. A Matlab implementation of the simplified version described in this section can be found in the ancillary files to \cite{Wil14c}.

Additional assumptions include:
\begin{itemize}
\item Probability of detection and survival are uniform, \textit{i.e.}\xspace, $P^{\mathrm{d}}(x)=P^{\mathrm{d}}$, $P^{\mathrm{s}}(x)=P^{\mathrm{s}}$
\item Target-originated measurements follow a linear-Gaussian model, $z_t = \mathbf{H} x_t + v_t$, where $v_t\sim\mathcal{N}\{v_t;0,\mathbf{R}\}$
\item Target dynamics follow a linear-Gaussian model, $x_t = \mathbf{F} x_{t-1} + w_t$, where $w_t\sim\mathcal{N}\{w_t;0,\mathbf{Q}\}$
\item Target birth intensity is a Gaussian mixture:
\[
\lambda^\mathrm{b}(x_t) = \sum_{k=1}^{n^b}\lambda^{b,k}\mathcal{N}\{x_t;\bar{x}^{b,k},\mathbf{P}^{b,k}\}
\]
\end{itemize}

\begin{figure}
\includegraphics{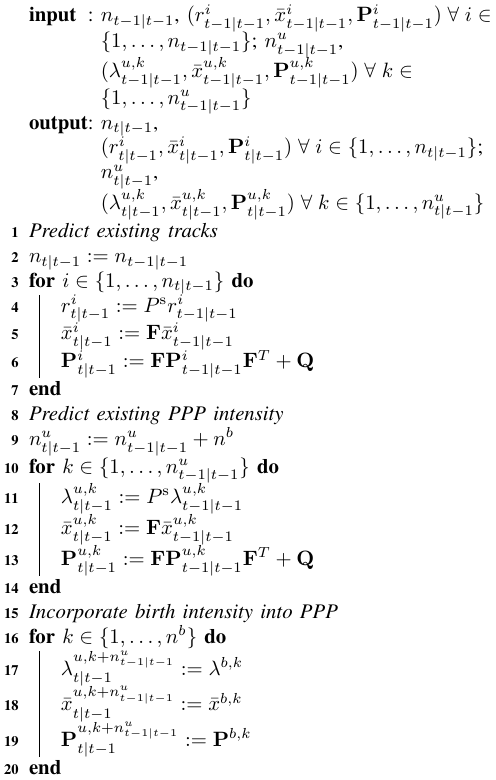}
\caption{Prediction algorithm.}
\label{fig:algPrediction}
\end{figure}

\begin{figure}
\ifCLASSOPTIONdraftcls
\includegraphics[width=3.5in]{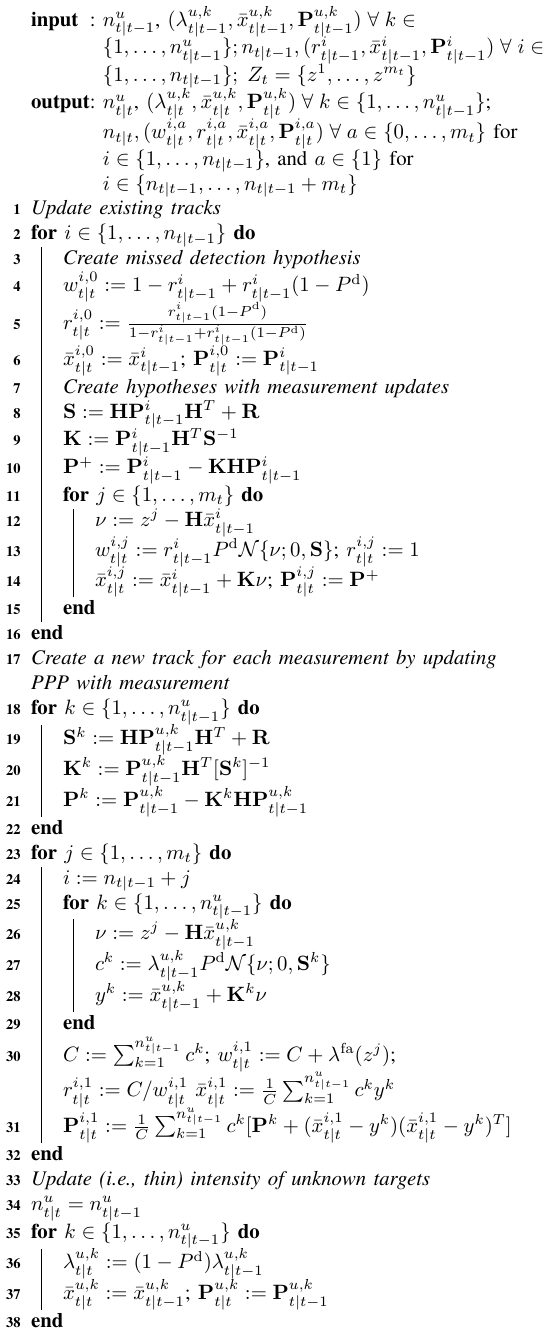}
\else
\includegraphics{figure8.pdf}
\fi
\caption{Component update algorithm.}
\label{fig:algUpdate}
\end{figure}

\begin{figure}
\includegraphics{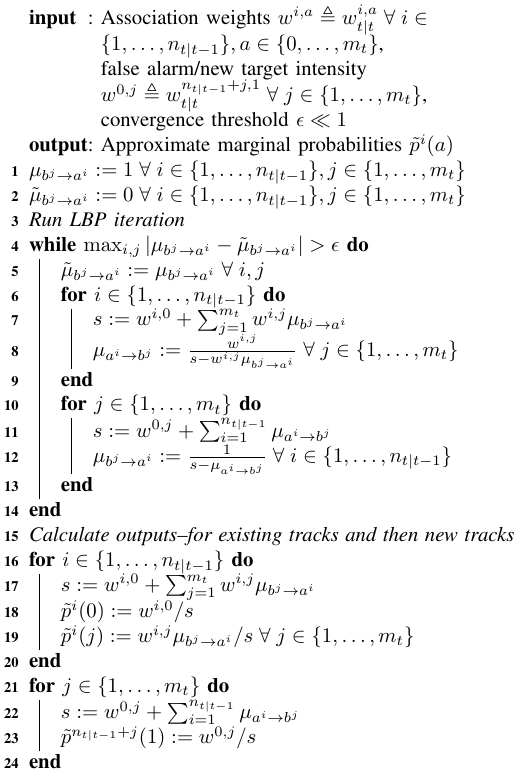}
\caption{LBP algorithm for approximation of marginal association probabilities.}
\label{fig:algLBP}
\end{figure}

\begin{figure}
\includegraphics{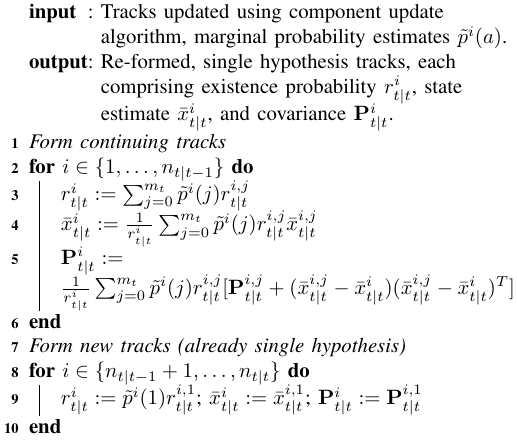}
\caption{TOMB/P algorithm for forming new tracks.}
\label{fig:algTOMMeMBerP}
\end{figure}

\begin{figure}
\includegraphics{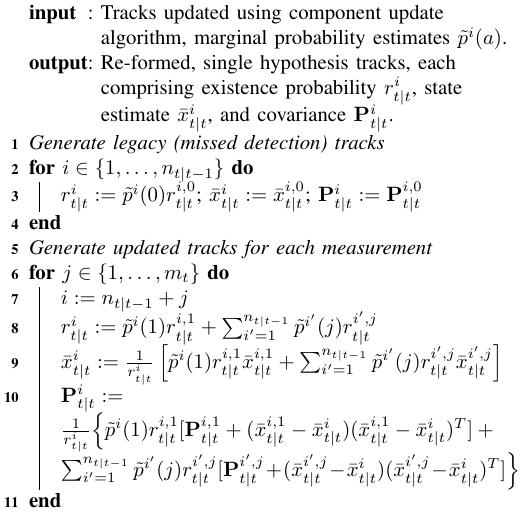}
\caption{MOMB/P algorithm for forming new tracks.}
\label{fig:algMOMMeMBerP}
\end{figure}

The state maintained by the algorithm is:
\begin{itemize}
\item A Gaussian mixture representation of the intensity of the PPP of unknown targets,
\[
\lambda^u_{t|t'}(x_t) = \sum_{k=1}^{n^u_{t|t'}}\lambda^{u,k}_{t|t'}\mathcal{N}\{x_t;\bar{x}_{t|t'}^{u,k},\mathbf{P}_{t|t'}^{u,k}\}
\]
\item $n_{t|t'}$ tracks (Bernoulli components), each comprised of a probability of existence $r^i_{t|t'}$, a mean $\bar{x}^{i}_{t|t'}$ and covariance $\mathbf{P}^i_{t|t'}$
\end{itemize}
The algorithm operates by executing the prediction step in Fig.~\ref{fig:algPrediction}, the update step in Fig.~\ref{fig:algUpdate}, LBP calculation of marginal association probabilities Fig.~\ref{fig:algLBP}, and then either the TOMB/P procedure for forming new tracks, Fig.~\ref{fig:algTOMMeMBerP}, or the MOMB/P procedure for forming new tracks, Fig.~\ref{fig:algMOMMeMBerP}. Steps for eliminating tracks with a low probability of existence and components in $\lambda^u_{t|t'}(x)$ with a low weight are not shown. Steps for extracting estimates are also not shown (TOMB/P outputs the mean estimate of each track with probability of existence greater than a threshold, while MOMB/P calculates the MAP cardinality estimate of the multi-Bernoulli distribution $\hat{n}$, and outputs the mean estimates of the $\hat{n}$ components with the highest probability of existence). Between update and re-forming of tracks, $(m_t+1)$ hypotheses exist for each pre-existing track $i$, parameterised by a hypothesis weight $w^{i,a}_{t|t}$, and hypothesis-conditioned probability of existence $r^{i,a}_{t|t}$, mean $\bar{x}^{i,a}_{t|t'}$ and covariance $\mathbf{P}^{i,a}_{t|t'}$. A single hypothesis exists for the new tracks, denoted by $a=1$ (the non-existence hypothesis in (\ref{eq:NewTargetNonExistWQ}) is accounted for without explicitly storing its weight, probability of existence, mean and covariance).

\ifCLASSOPTIONdraftcls
\else
%\IEEEtriggercmd{\enlargethispage{-0.2in}}
%\IEEEtriggeratref{20}
\fi

{\small{\bibliographystyle{IEEEtran}
\bibliography{../../IEEEabrv,../../Bibliography}}}

\ifCLASSOPTIONdraftcls
\else
\vfill
\fi

\begin{IEEEbiography}[{\includegraphics[width=1in,height=1.25in,clip,keepaspectratio]{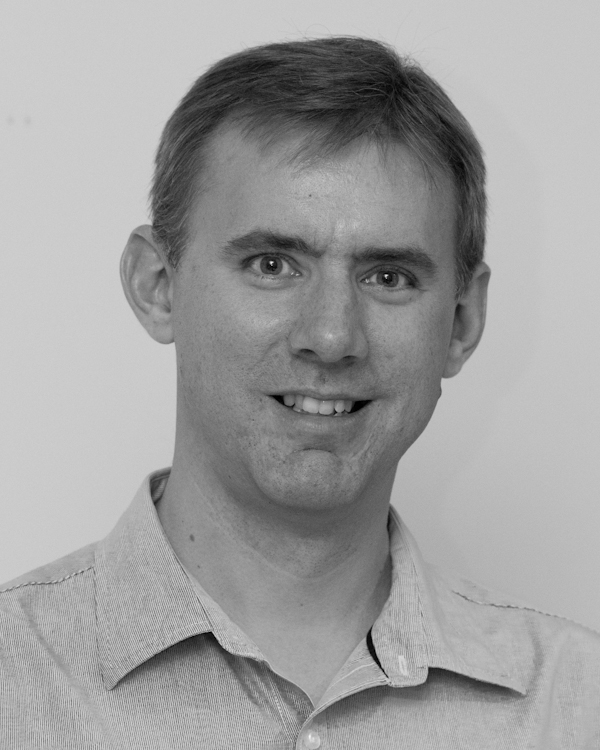}}]{Jason L.\ Williams} (S'01--M'07) received degrees of BE(Electronics)/BInfTech from Queensland University of Technology in 1999, MSEE from the United States Air Force Institute of Technology in 2003, and PhD from Massachusetts Institute of Technology in 2007. 

He worked for several years as an engineering officer in the Royal Australian Air Force, before joining Australia's Defence Science and Technology Organisation in 2007. He is also an adjunct senior lecturer at the University of Adelaide. His research interests include target tracking, sensor resource management, Markov random fields and convex optimisation. 
\end{IEEEbiography}

\end{document}